\documentclass[12pt]{iopart}
\usepackage{indentfirst}
\usepackage{graphicx,color} 
\usepackage{iopams}

\expandafter\let\csname equation*\endcsname\relax

\expandafter\let\csname endequation*\endcsname\relax

\usepackage{amsmath}
\usepackage[caption=false]{subfig}

\usepackage[latin9]{inputenc}
\newcommand{\avg}[1]{\left< #1 \right>} 

\definecolor{Blue}{rgb}{0.0,0.0,1}
\definecolor{Red}{rgb}{1,0.0,0.0}
\definecolor{Green}{rgb}{0,0.5,0.0}

\begin{document}

\title[Effective temperatures for single particle system under dichotomous noise]{Effective temperatures for single particle system under dichotomous noise} 

\author{Jo\~ao R. Medeiros and S\'ilvio M. Duarte~Queir\'os}
\address{Centro Brasileiro de Pesquisas F\'isicas, Rua Dr Xavier Sigaud 150, 22290-180 Urca, Rio de Janeiro - RJ, Brazil}
\ead{joaom@cbpf.br , sdqueiro@cbpf.br}
\vspace{10pt}
\begin{indented}
\item[] \today
\end{indented}

\begin{abstract} 
Three different definitions of effective temperature -- $\mathcal{T}_{\rm k}$, $\mathcal{T}_{\rm i}$ and $\mathcal{T}_{\rm r}$  related to kinetic theory, system entropy and response theory, respectively -- are applied in the description of a non-equilibrium generalised massive Langevin model in contact with dichotomous noise. The differences between the definitions of $\mathcal{T}$ naturally wade out as the reservoir reaches its white-noise limit, approaching Gaussian features. The same framework is employed in its overdamped version as well, showing the loss of inertial contributions to the dynamics of the system also makes the three mentioned approaches for effective temperature equivalent.
\end{abstract}

\maketitle

\section{Introduction}

\par The present work is an effort to answer the following question: is it possible to define an effective temperature in the context of an out of equilibrium single particle 1-D system?
Even though a single effective measure may not be able to map all temperature properties that emerge in thermal equilibrium, it is possible to investigate how each specific aspect attributed to temperature behaves in non-equilibrium. Correspondingly, it is feasible to define respective effective measures, resulting in a set of effective temperatures, which are bound to converge to the single canonical temperature as the system approaches thermal equilibrium. 

Since convergence is a necessary condition for equilibrium, to address it in the context of a stripped down model is to delve into the issue of quantifying the approach to equilibrium in fundamental terms, shedding light on the essential ingredients that differentiate the features of temperature.

\par The concept of temperature is paramount in the thermostatistical description of a system, let us recall it is explicitly mentioned in the statements of the zeroth and third laws of Thermodynamics,  as well as in the definition of thermal reservoir~\cite{huang,landsberg}. Although intimately related to equilibrium, the notion of temperature has been extended to non-equilibrium instances as well: from glassy~\cite{glassy} to granular~\cite{granular} passing through chaotic nonlinear systems~\cite{temp-maps}, active matter~\cite{fodor} alongside other problems~\cite{parisi-casas,cugliandolo-vulpiani}.
In all of them, one verifies the system is  either isolated or in contact with Gaussian (coloured) thermal-like sources;  hence, their stochastic dynamical equations are associated with Fokker-Plank Equations~\cite{risken}.

More intricate systems have shown that interactions with sources of energy other than heat are not sufficiently well described by Gaussian noises --- or even the mapping into it~\cite{mexicanos} --- especially at the level of the Large Deviation behaviour~\cite{jrm-smdq-physa}. Considering this fact, Generalised Langevin equations which do not translate into Fokker-Planck Equations (by generating infinite non-zero terms in the Kramers-Moyal expansion following Pawula's Theorem~\cite{risken}) have ensued as powerful tools for the statistical description of such data~\cite{sevens}. 

Even though systems in contact with non-Gaussian reservoirs can be far off the usual equilibrium picture, we assert that
it should be possible to come up with an equivalent thermal description as long as one is able to bring forth an adequate definition of effective temperature~\cite{iti}.

In this manuscript, we study the impact of different definitions of effective temperature by considering a paradigmatic (massive) damped system in contact with dichotomous noise, also known as 2-state Brownian or telegraphic noise~\cite{caceres}. Comparing three different approaches we provide a definition of a metric by which one quantifies the difference between intensive, kinetic and response aspects of temperature in non-equilibrium for the present model. Our framework is easily adaptable to other reservoirs, given that the
two leading ingredients of standard non-Gaussian reservoirs are present in our model: colour and high-order cumulants~\cite{br-jp, br-jp-2}. Besides, dichotomous noise can be used to analytically represent systems exhibiting stochastic tug-of-war mechanism, as in problems of intracellular bidirectional transport on cytoskeletal filaments~\cite{muller}, transport properties in amorphous materials~\cite{pfister}, calcium ions motion in blood plasma~\cite{gitterman}, active matter~\cite{run-tumble}, fluctuations of photovoltaics in power grids~\cite{matthias}, particular types of ratchets~\cite{kolomeisky}, among other problems. 

At the end of our survey, we establish analytical one-to-one relationships between effective measures $\mathcal{T}_{\rm k}$, $\mathcal{T}_{\rm i}$ and $\mathcal{T}_{\rm r}$. These relations are indicative of the distinct ways with which the corresponding properties of kinetic energy, system entropy and response influence each other and balance out their mutual differences in the white-noise limit, where the system approaches thermal equilibrium.

\bigskip

\section{The underdamped case}

We have considered a non-equilibrium system with mass $m$, and position $x$, ruled by the following model equation:
\begin{equation}
m\frac{d^{2}x\left(t\right)}{dt^{2}}=-\gamma\,\frac{dx\left(t\right)}{dt}-k\,x\left(t\right)+\zeta_t, \quad v\left(t\right) \equiv \frac{dx\left(t\right)}{dt}
\label{modelo}
\end{equation}

The parameter $\gamma$ relates to some type  of friction the system is subjected to, and $\zeta$ is the stochastic force describing the interaction between the particle and the dichotomous reservoir for which we use the Stratonovich interpretation. The confinement is established by the harmonic potential, $k\,x^{2}/2$, which can represent the features of the system or else the action of an optical tweezer --- the behaviour of which is known to be very close to harmonicity~\cite{nbx97}. 

Regarding $\zeta_t $ it assumes two symmetric values, $\zeta_t \in \left\{ -a,a\right\} $, hopping between them with the same transition rate, $\mu$, allowing for a bimodal distribution, 
$f\left(\zeta\right)=\frac{1}{2}\,\delta\left(\zeta+a\right)+\frac{1}{2}\,\delta\left(\zeta-a\right) $
and a coloured correlation function with frequency, $\alpha=2\,\mu$:
\begin{equation}
\left\langle \left\langle \zeta\left(t_{1}\right)\,\,\zeta\left(t_{2}\right)\right\rangle \right\rangle  = a^{2}\,\mathrm{e}^{-\alpha\,\left\vert t_{1}-t_{2}\right\vert }
\label{telegraphcor}
\end{equation}

See appendix A for details on the dichotomous noise regarding its more general definition and numerical simulation.
Note that $\left\langle \left\langle \ldots\right\rangle \right\rangle $ stands for cumulants, the single brackets $\left\langle \ldots\right\rangle $ will be used for averages over samples.
Differently to Gaussian cases, higher-order cumulants $\left\langle \left\langle \zeta\left(t_{1}\right)\,\,\zeta\left(t_{2}\right) \ldots \zeta\left(t_{n}\right) \right\rangle \right\rangle$ do not vanish and thus the L\'evy-It\^o theorem on the decomposition of measure cannot be applied~\cite{applebaum}.  Still, dichotomous noise can be deemed a work reservoir, because it acts on the massive system by either pulling or pushing it during random periods of time. 

From Eq.~(\ref{modelo}), it is possible to identify the following features: \emph{i)} Since $\zeta$ is coloured and the dissipation term, $-\gamma\,v\left(t\right)$, has no kernel, the dichotomous work reservoir is of external class~\cite{kubo,vankampen}; 
\emph{ii)} the particle cannot explore its full phase space, except in the limit $\alpha\rightarrow\infty$, which was analytically proven in Ref.~\cite{JRM-SMDQ-2015} alongside other statistical features of $\left(x,v\right)$ by means of an averaging procedure in Laplace-Fourier space.

\subsection{Kinetic Temperature}
\bigskip

The long-term time averaged energy fluxes of the Langevin particle in contact with Gaussian reservoir have  the following text-book formulae:
\begin{gather}
\it{j}_{{\rm dis}}^{\left({\rm Langevin}\right)}    \equiv  \lim _{t \rightarrow \infty} \frac{1}{t} \int_{0}^{t}\left[ -\gamma \,v\left( s\right) \right] \,v\left( s\right) 
\,ds
\label{Jdis}
\\
\it{j}_{{\rm inj}}^{\left({\rm Langevin}\right)}    \equiv \lim _{t \rightarrow \infty} \frac{1}{t} \int_{0}^{t }\eta \left( s\right) \,v\left( s\right) \,ds 
\label{Jinj}
\end{gather}
where $\eta(t)$ represents the Gaussian reservoir term in the Langevin model equation (Eq.(\ref{modelo}) with $\eta(t)$ in place of $\zeta(t)$). The terms  $\it{j}_{{\rm dis}}$ an $\it{j}_{{\rm inj}}$ are respectively the dissipated heat over time and the injected heat over time. The calculations yield;
\begin{equation}
\it{j}_{{\rm inj}}^{\left({\rm Langevin}\right)}  
= -  \it{j}_{{\rm dis}}^{\left({\rm Langevin}\right)}  
=  \frac{\gamma}{m} T  
\label{JfromT}
\end{equation}
%


For the generalized Langevin particle under the dichotomous reservoir, we can make the same calculations (now with $\zeta(t)$ in place of $\eta(t)$). Notice when calculating averages we use the Laplace-Fourier method,  see appendix B for details. We obtain;  
\begin{equation}
 \it{j}_{{\rm inj}}^{\left({\rm dichotomous}\right)}  
= -  \it{j}_{{\rm dis}}^{\left({\rm dichotomous}\right)} 
= \frac{a^2 \alpha  }{  (k+\alpha  (\gamma +\alpha  m))} 
\end{equation}
Considering kinetic theory~\cite{huang}\cite{JRM-SMDQ-2015} we have the first prospective for a definition of effective temperature. Let us define $\mathcal{T}_{\rm k} $ as the kinetic temperature;
\begin{equation}
\mathcal{T}_{\rm k} \equiv m \, \left\langle v^{2}\right\rangle = \frac{a^2 \alpha  m}{\gamma  (k+\alpha  (\gamma +\alpha  m))}
\label{marconi}
\end{equation}
Thence, we can write;
\begin{equation}
 \it{j}_{{\rm inj}}^{\left({\rm dichotomous}\right)}  
= -  \it{j}_{{\rm dis}}^{\left({\rm  dichotomous}\right)} 
= \frac{\gamma}{m} \mathcal{T}_{\rm k}  
\end{equation}

Recovering the functional form of Eq.(\ref{JfromT}). This result was originally presented in section IV of Ref. \cite{JRM-SMDQ-2015} and section II, subsection C of Ref. \cite{heatflux}. Note that the total energy of the system reads:

\begin{equation}
E =\frac{m}{2}\left\langle v^{2}\right\rangle + \frac{k}{2}\left\langle x^{2}\right\rangle  = \frac{a^2 (\gamma +2 \alpha  m)}{2 \gamma  \left(\alpha  \gamma +k+\alpha ^2 m\right)}
\end{equation}

%
%
The analogous of  equipartition of energy --- i.e., $E = T$ given $m\left\langle v^{2}\right\rangle/2 = T/2$ and $k\left\langle x^{2}\right\rangle/2 = T/2$ --- is not verified though. It actually reads:
\begin{equation}
E =\mathcal{T}_{\rm k} \left(1+ \frac{\gamma }{2 \, m \, \alpha} \right)
\end{equation}
which shows an explicit dependence on the relaxation, $\gamma$, and colour, $\alpha$, scales, respectively. When $\gamma << \alpha$, the standard equipartition relation is recovered:
\begin{equation}
E = \mathcal{T}_{\rm k}
\end{equation}
Note that the results contained in this section are valid for any sort of noise that generates an auto-correlation function of the form displayed in Eq.~(\ref{telegraphcor}).

\subsection{Intensive Temperature}

Taking into consideration the postulates of Thermodynamics, ~\cite{landsberg}, we set forth the definition of intensive temperature, as it corresponds to the conjugate intensive field of the system entropy $S$: 
\begin{equation}
\frac{1}{\mathcal{T}_{\rm i}} \equiv \frac{\partial S}{\partial E}
\label{tempintdef}
\end{equation}
where $S = -  \int\int p(x,v) \, \log p(x,v) \, dx \, dv$ (we have set \mbox{$k_B$ = 1}).

Having the Langevin case as benchmark, wherein $T$ plays the part of both $\mathcal{T}_{\rm i}$ and $\mathcal{T}_{\rm k}$, we have focused on understanding to what extent the two measures differ one another in the dichotomous reservoir case. Our framework is one that emulates the traditional canonical ensemble by fixing the effective temperature $\mathcal{T}_{\rm k}$ instead of the equilibrium temperature $T$. This proposal is called the quasi-canonical ensemble.

Assuming the mass $m$ and the mechanical parameters $\gamma $ and $k$ of the system are kept constant, the partial derivative in Eq.~(\ref{tempintdef}) turns into an absolute derivative conditioned to $\mathcal{T}_{\rm k}$, but with different combinations of reservoir amplitude $a$ and reservoir colour $\alpha$. Manipulating Eq.~(\ref{tempintdef}),  we have obtained (see appendix C for the step by step derivation):
\begin{equation}
\frac{1}{\mathcal{T}_{\rm i}}  = \frac{dS}{dE}  \bigg\vert_{\mathcal{T}_{\rm k}} =   \frac{\partial S}{\partial E}   \bigg\vert_{\alpha} - \frac{2 m \, \alpha^2 }{\gamma  \, \mathcal{T}_{\rm k}} \, \frac{\partial S}{\partial \alpha }  \bigg\vert_{E} 
\label{tibimodal}
\end{equation}

\begin{figure}[tbh]
\begin{minipage}{.5\linewidth}
\centering
\subfloat[]{\includegraphics[scale=.6]{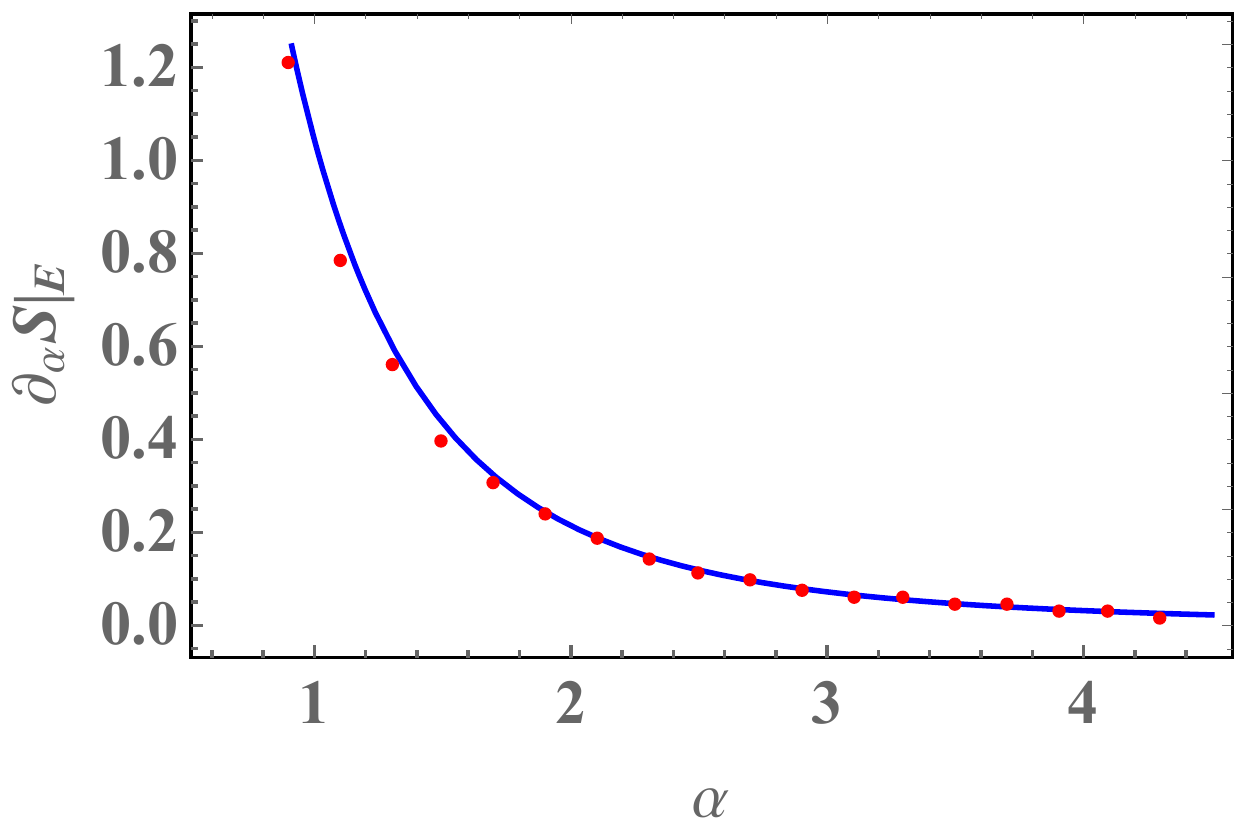}}
\end{minipage}%
\begin{minipage}{.5\linewidth}
\centering
\subfloat[]{\includegraphics[scale=.6]{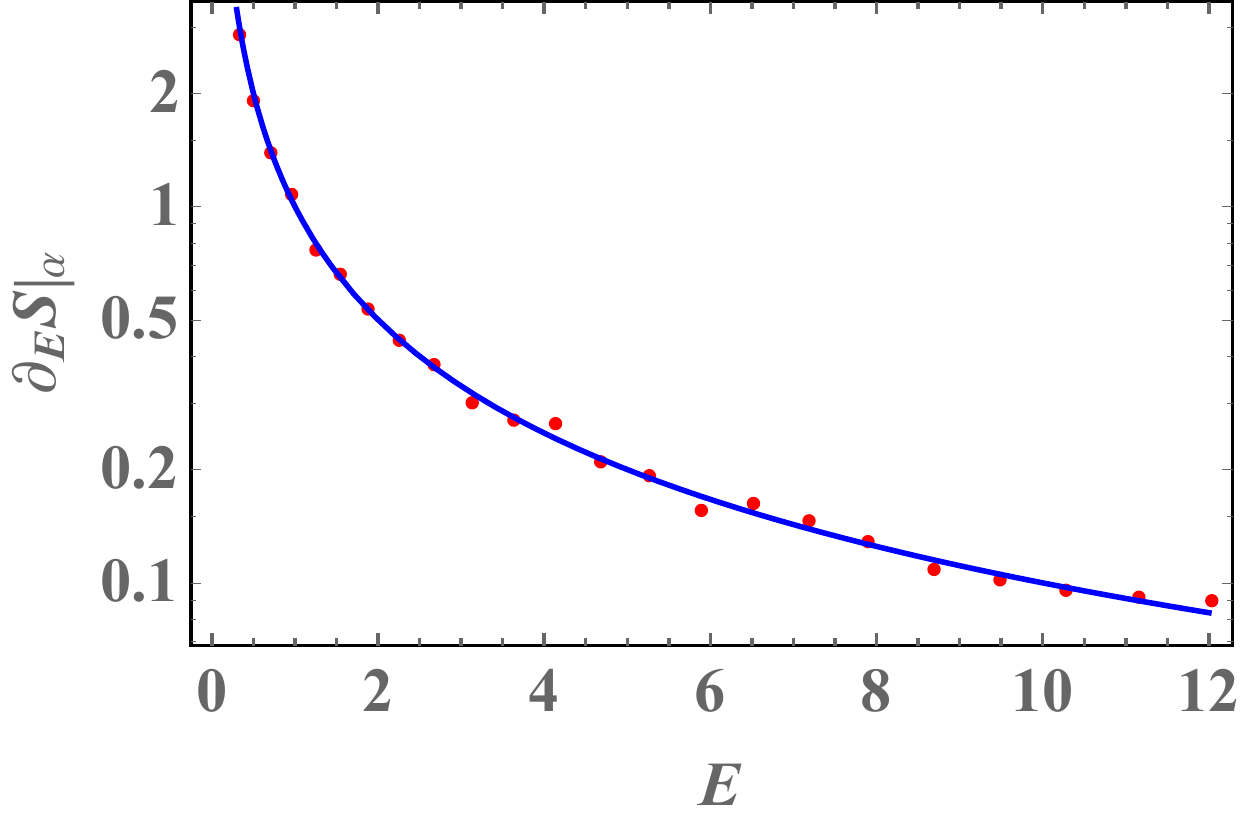}}
\end{minipage}\par\medskip
\centering
\begin{minipage}{.5\linewidth}
\centering
\subfloat[]{\includegraphics[scale=.6]{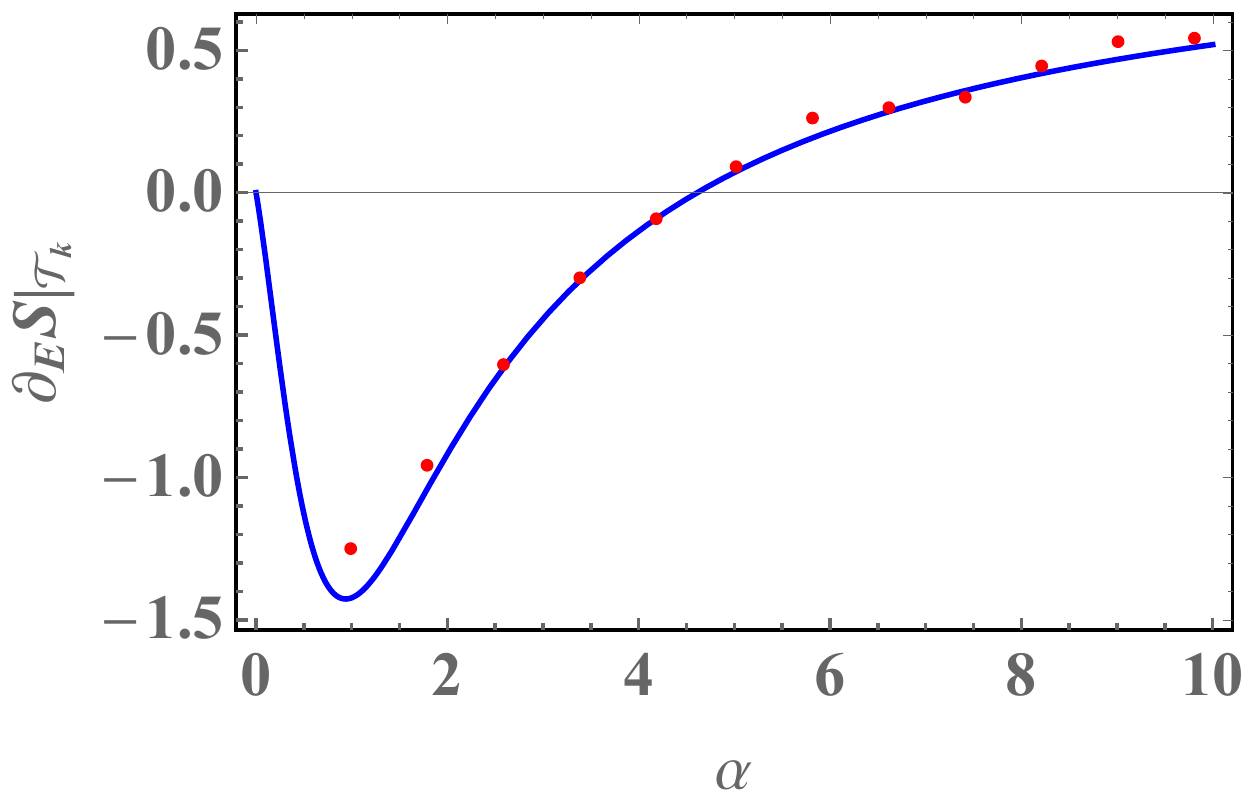}}
\end{minipage}%
\begin{minipage}{.5\linewidth}
\centering
\subfloat[]{\includegraphics[scale=.6]{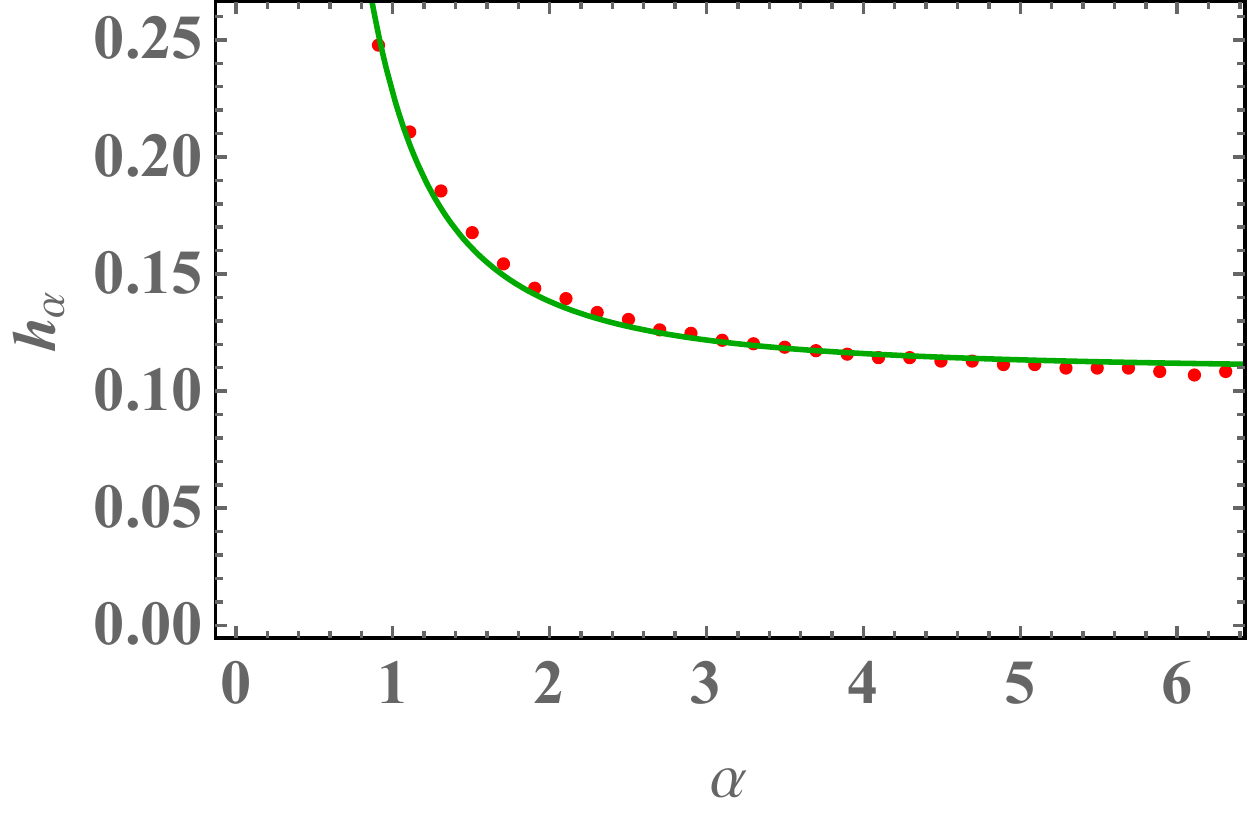}}
\end{minipage}\par\medskip
\caption{
Panel a): ${\partial}_{ \alpha } S \Large\vert_{E  = 3/10}$ versus $\alpha $. Panel b): ${\partial }_{E} S \Large\vert_{\alpha = 14/10}$  versus $E$. Panel c) ${\partial }_{E} S \Large\vert_{\mathcal{T}_{\rm k} = 1}$ versus $\alpha$. Blue lines correspond to values for the partial derivatives in Eq.~(\ref{tibimodal}) obtained using the analytical proposal Eq.~(\ref{entropyansatz}),  whereas the red dots have been obtained from the numerical implementation of Eq.~(\ref{modelo}). Panel d) exhibits the values for $h_\alpha$ from Eq.~(\ref{entropyansatz}), red dots are obtained by adjusting the ansatz to simulations of the model, and green line corresponds the formula $h_{\alpha} = \left( h^{-1} + \frac{h^{\prime} }{\alpha^2} \right)^{-1}$. The mechanical parameters were set to $m = k = \gamma = 1$.
}
\label{partialderivatives} 
\end{figure}

In the expression above, we identify $\frac{m}{\gamma  \, \mathcal{T}_{\rm k}} = j ^{-1}$ with $j$ being the long-term average power or energy flux per unit time, see Eq.~(\ref{JfromT}) for the energy fluxes. The behaviour of the partial derivatives is depicted in Fig.~\ref{partialderivatives}.

It is known the steady state distribution $p(x,v)$ of Eq.~(\ref{modelo}) does not have a closed form~\cite{caceres,czernik}, thus we have resorted to its numerical implementation --- within a sufficiently large range of the  reservoir parameters ($\alpha$, $a$). Then, the total system entropy, $S \equiv - \int  p(x,v) Log \big( p(x,v) \big) dx dv $ is computed where the portion of the dynamics corresponding to the transient is jettisoned.

\begin{figure}[tbh]
\begin{centering}
\includegraphics[width=0.5\columnwidth]{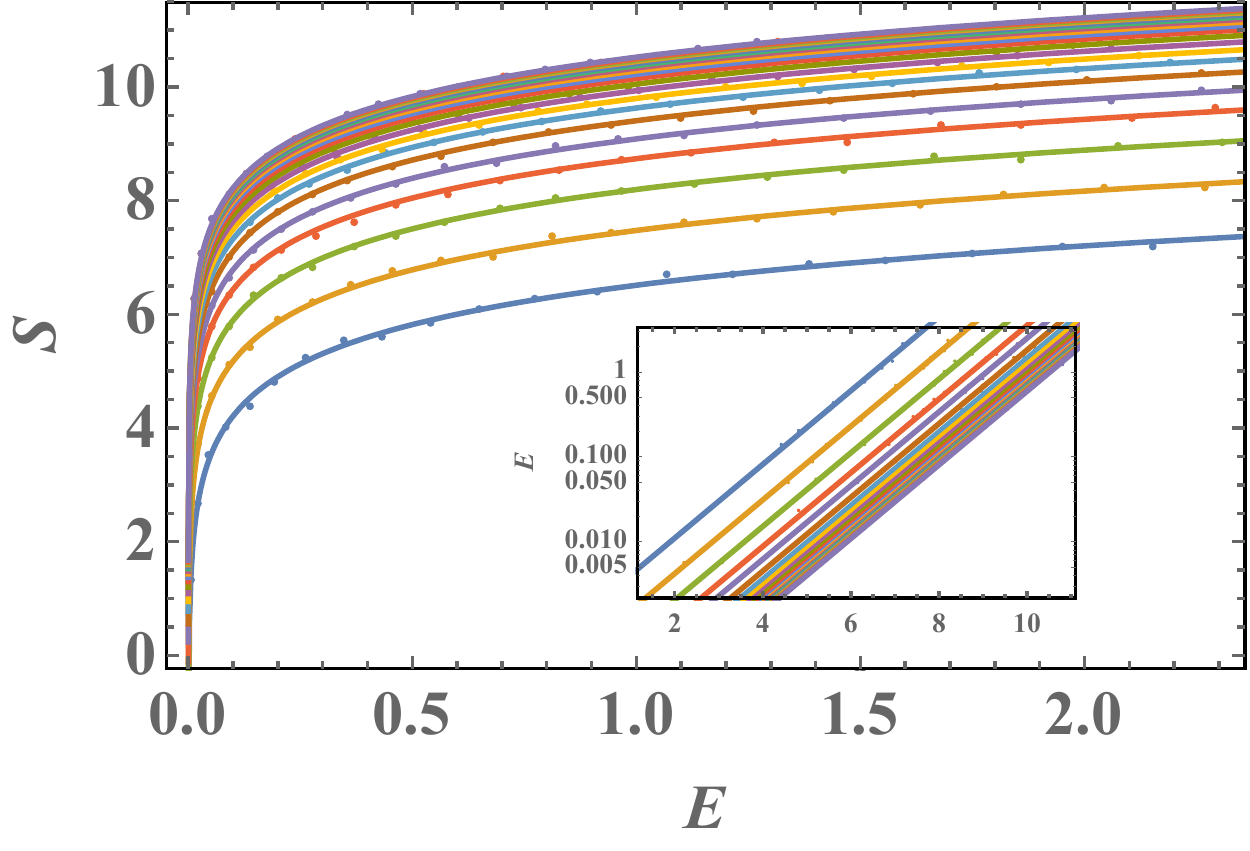}
\includegraphics[width=0.5\columnwidth]{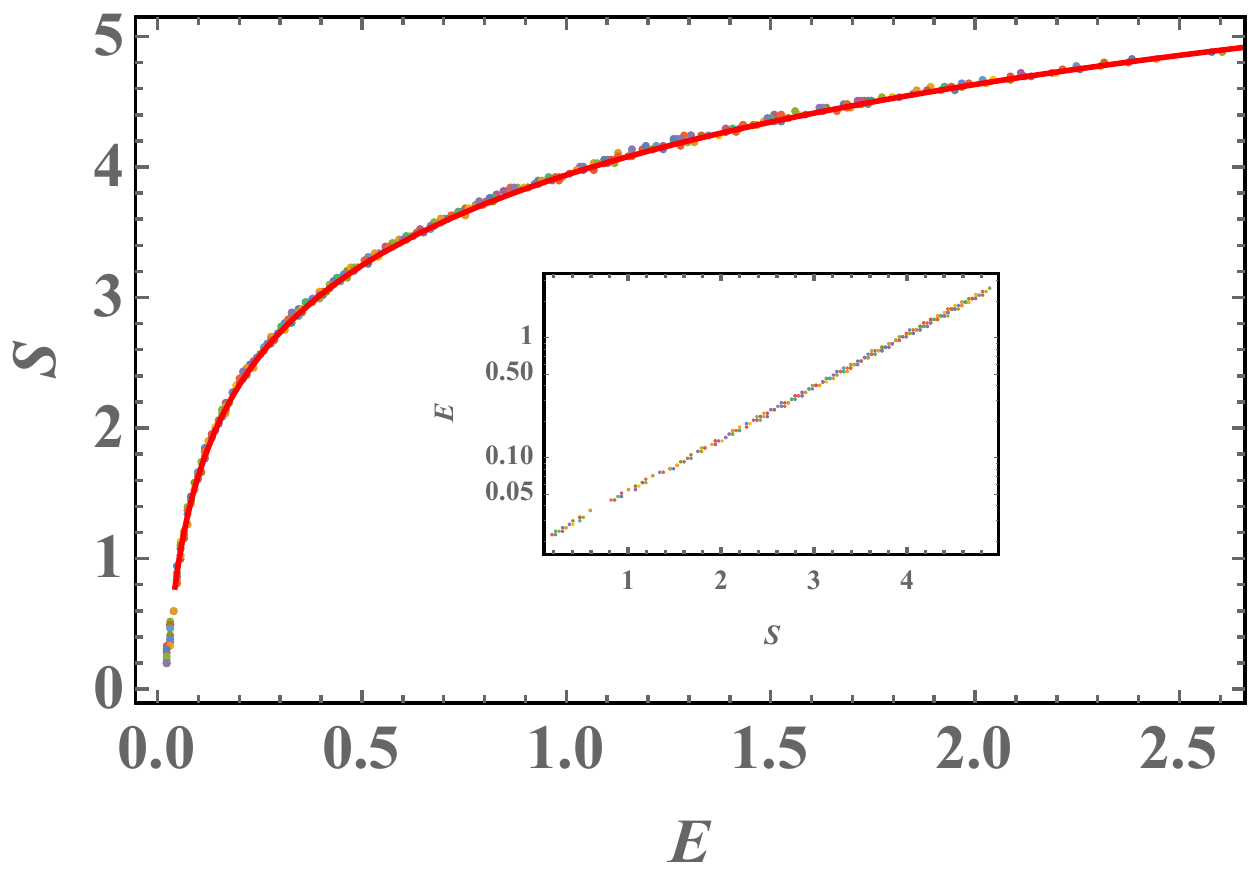}
\end{centering}
\caption{Entropy versus Energy, different colors correspond to different values of $\alpha $  with $\mathcal{T}_{\rm k} = 0.710\ldots $ and $\gamma = m = k = 1$. Left panel: starting with the blue line where $\alpha = 1/10$, each next line going up is a set of simulations and analytical predictions with bigger $\alpha$. Note that each point in the plot has a unique set of values for colour and amplitude, respectively $\alpha, a \in \{1/10 ,  2/10 , 3/10 ..., 46/10 \}$ . Right panel: convergence is attained, all different values of $\alpha$ generate roughly the same values of entropy and energy $\alpha \in \{3 ,  32/10 , 34/10 ..., 108/10 \}$. The inset in each panel shows the validity of Eq.~(\ref{entropyansatz}).}
\label{SXEE} 
\end{figure}

Using the results plotted in Fig.~\ref{SXEE}, we assert the entropy is very well described by the ansatz:
\begin{equation}
S= \log ( h_{\alpha} \, E )
\label{entropyansatz}
\end{equation}
with $h_{\alpha} = \left( h^{-1} + \frac{h^{\prime} }{\alpha^2} \right)^{-1}$, where $h$ and $h'$ are functions of the mechanical parameters and the noise amplitude, estimated numerically for a set $\{ m, k , \gamma , a\}$  by adjusting the ansatz Eq.~(\ref{entropyansatz}). This approach is implemented as a way to investigate how the persistence scale of the reservoir --- which is expressed by the parameter $\alpha$ --- affects the mutual relation between entropy, energy and the intensive effective temperature, therefore analytical expressions for $h(m, k , \gamma, a)$ and $h'(m, k , \gamma, a)$ are not worth pursuing given the current framework. Nevertheless, the logarithmic dependence in Eq.~(\ref{entropyansatz}) can be proven analytically in the overdamped regime, as will be shown in the next section. A procedure to retrieve expressions for the explicit dependence of $h$ and $h^\prime $ on the dynamical parameters in that case is also presented.

From Eq.~(\ref{entropyansatz}), we have obtained:
\begin{equation}
\frac{1}{\mathcal{T}_{\rm i}} = \frac{1}{\mathcal{T}_{\rm k}} \,
\left( \frac{2\, m \,\alpha}{\gamma + 2\, m \, \alpha} - \frac{4\, m \, h^{\prime} \, \alpha ^2}{\gamma \, \alpha  \left(\alpha ^2 / h +  h^{\prime}\right)} \right)
\label{Ti-Tv}
\end{equation}

For $\alpha^{*} =  \frac{\sqrt{\gamma ^2 h h'+4 m^2 \left(h'\right)^2}+2 m h'}{\gamma  h},$ 
$\mathcal{T}_{\rm i}$ becomes undefined and changes sign. The negative region $\alpha<\alpha^{*} $ corresponds to a large transition time between the values of the dichotomous noise. In that case diversity of states is dramatically diminished and a quasi-deterministic solution to Eq.~(\ref{modelo}) can be implemented~\cite{SMDQ-FR}. From a statistical mechanics standpoint, the model under that premise resembles the text-book case of the two-level energy system which yields negative values and a divergent critical value for temperature as well. The critical point $\alpha =\alpha^{*} $ may be interpreted as the reservoir setup at which the system becomes the most detached from thermal equilibrium in regards to the relationship between entropy and energy. Note that $\alpha =\alpha^{*} $ corresponds to ${\partial }_{E} S \Large\vert_{\mathcal{T}_{\rm k} } = 0 $, the singular point of parameter space where a variation on the value of the system energy does not produce a corresponding change in the system entropy, with both quantities becoming unrelated at the transition between the negative and positive $\mathcal{T}_{\rm i}$ regions. As $\alpha$ goes to infinity --- and the noise amplitude, $a$, changes so that $\mathcal{T}_{\rm k}$ is kept constant --- one obtains a situation that mimics the Gaussian reservoir case and $\mathcal{T}_{\rm i} = \mathcal{T}_{\rm k} = T$, as expected. It is also relevant to mention that many aspects of the steady-state distribution $p(x,v)$ undergo some form of  critical behaviour determined by parameter $\alpha$, such as the kurtosis and skewness \cite{JRM-SMDQ-2015} of the distribution, hence the critical behaviour displayed by $\mathcal{T}_{\rm i}$ at $\alpha^{*}$ is not unprecedented. 

The results for intensive temperature herein displayed are specific to the presently discussed particle under dichotomous noise, given that they reflect  entropic aspects related to the exotic stationary probability distributions generated by this model, see Refs. \cite{JRM-SMDQ-2015}, \cite{SMDQ-FR}.
\subsection{Response Temperature}
In our third approach for an effective temperature we resort to the study of the correlations in the system and their relation to response functions, in short, we investigate the possibility of establishing an alternative relation that plays the part of an effective Fluctuation-Dissipation Theorem from which an effective measure of temperature might be derived.

That picture recalls Kubo's response theory~\cite{kubo}, where the investigation of time-dependent correlations in the system leads to the Fluctuation-Dissipation Theorem. Even though such approach was originally derived from assumptions related to thermal equilibrium, it has been successfully considered in several out of equilibrium systems as well~\cite{cugliandolo-vulpiani,fodor} ~\cite{iti}. 

Along the same lines, we have considered $ R (t,t^\prime)$ as the response  function of the system --- after it has attained the steady state ---  to the action of a small constant field~\cite{risken}, $f(t) \equiv f \, \Theta \left(t-t'\right)$,  at instant $t^\prime$:
\begin{equation}
R (t,t^\prime) \equiv \frac{\delta \left\langle x \left(t\right)\right\rangle}{ \delta f (t^\prime) }
\label{reponse}  
\end{equation}
The integral counterpart to the response function, the susceptibility, reads:
\begin{equation}
\chi (t, t^\prime) \equiv \int_{ t^\prime} ^t  R (t,s) ds
\label{susceptibility} 
\end{equation}
Concomitantly, standard response theory bridges the response and the correlation functions in the thermal equilibrium Gaussian case, producing the fluctuation-dissipation theorem:
\begin{equation}
\frac{ \partial C_x \left(t,t^\prime \right)}{\partial t^\prime } \equiv \frac{ \partial \left\langle x\left(t\right)x \left(t^\prime \right) \right\rangle}{\partial t^\prime } = T\, R \left(t,t^\prime \right)
\label{eqresponse}
\end{equation}
We shall investigate the position auto-correlation displayed by the massive particle under dichotomous noise. The result is obtained by Fourier-Laplace transforming the dynamical equation and complex integrating the result (see appendix B). For Eq.~(\ref{modelo}), the steady-state correlation is given by:
\begin{equation}
\left\langle x(t)x(t+s) \right\rangle = {\rm const} \,\, \left[ e^{-\frac{\gamma }{2 m}\, s} \frac{1}{k}\left(\frac{\gamma  \left( 2\,k\,m+\Lambda \right) }{2 \,m \, \Omega \, \Lambda }\sin \left(  \Omega \, s \right)+\cos \left(   \Omega s \right)\right) -\frac{\gamma }{\alpha \, \Lambda } e^{- \alpha \,  s}  \right] ,\qquad \left( s \ge 0 \right)
\label{eqeqcorrelation}
\end{equation}
with $\Omega \equiv \sqrt{4 k m-\gamma ^2}/\left(2\,m\right)$ and $ \Lambda \equiv m \left(k+\alpha ^2 m\right)-\gamma ^2$. 

As we have mentioned, the reservoir in this problem is of external class since the dissipation and the noise spectra mismatch, leading to the violation of Kubo's Theorem. Nevertheless, with the purpose of understanding effective temperature behaviour in the context of response theory, we establish an alternate version of the Fluctuation-Dissipation Theorem that is verified in the current scenario.

Referring to the previous section where we have found the convergence of $\mathcal{T}_{\rm k}$ and $\mathcal{T}_{\rm i}$ at the white noise limit of $\alpha \rightarrow \infty$ while keeping $\mathcal{T}_{\rm k}$ constant, we shall investigate how the same limit affects the position auto-correlation expression Eq.~(\ref{eqeqcorrelation}).

In fact, the white-noise limit produces  the convergence of the factor indicated by $\rm {const}$ to  $\mathcal{T}_{\rm k}$ and the full expression for the auto-correlation becomes a modified form of the Langevin linear susceptibility. We designate the factor, $\rm {const}$, as the response temperature $ \mathcal{T}_{\rm r}$ and remaining expression of the position auto-correlation as the effective susceptibility $\chi_{{\rm eff}}$, following a close path to that conveyed in Ref.~\cite{fodor}, where, in a complementary approach to ours, an expression for effective correlation is derived from the susceptibility;
\begin{equation}
C_x(t,t^\prime)= \mathcal{T}_{\rm r} \, \chi_{{\rm eff}} (t, t^\prime).
\label{Tr}
\end{equation}
the derivation of both $ \mathcal{T}_{\rm r}$ and $\chi_{{\rm eff}}$ is presented. In conclusion, Eq.~(\ref{Tr}) works as the effective fluctuation-dissipation relation Eq.~(\ref{eqresponse}).

As made for $\mathcal{T}_{\rm i}$ in Eq.~(\ref{Ti-Tv}), the definition of $\mathcal{T}_{\rm r}$ is expressed in relation to constant kinetic temperature:%
\begin{equation}
{\rm const } = \mathcal{T}_{\rm r} \equiv  \mathcal{T}_{\rm k} \, \frac{\left(k+\alpha ^2 m\right)-\gamma ^2/m}{k+\alpha  \left(\alpha \, m-\gamma \right)} . %
\label{Tr-Tv}
\end{equation}

Note that all results related to the response aspects of the system are valid for any noise that generates a correlation function of the form given by Eq.~(\ref{telegraphcor}).

\begin{figure}[ht]
\begin{centering}
\includegraphics[width=0.5\columnwidth]{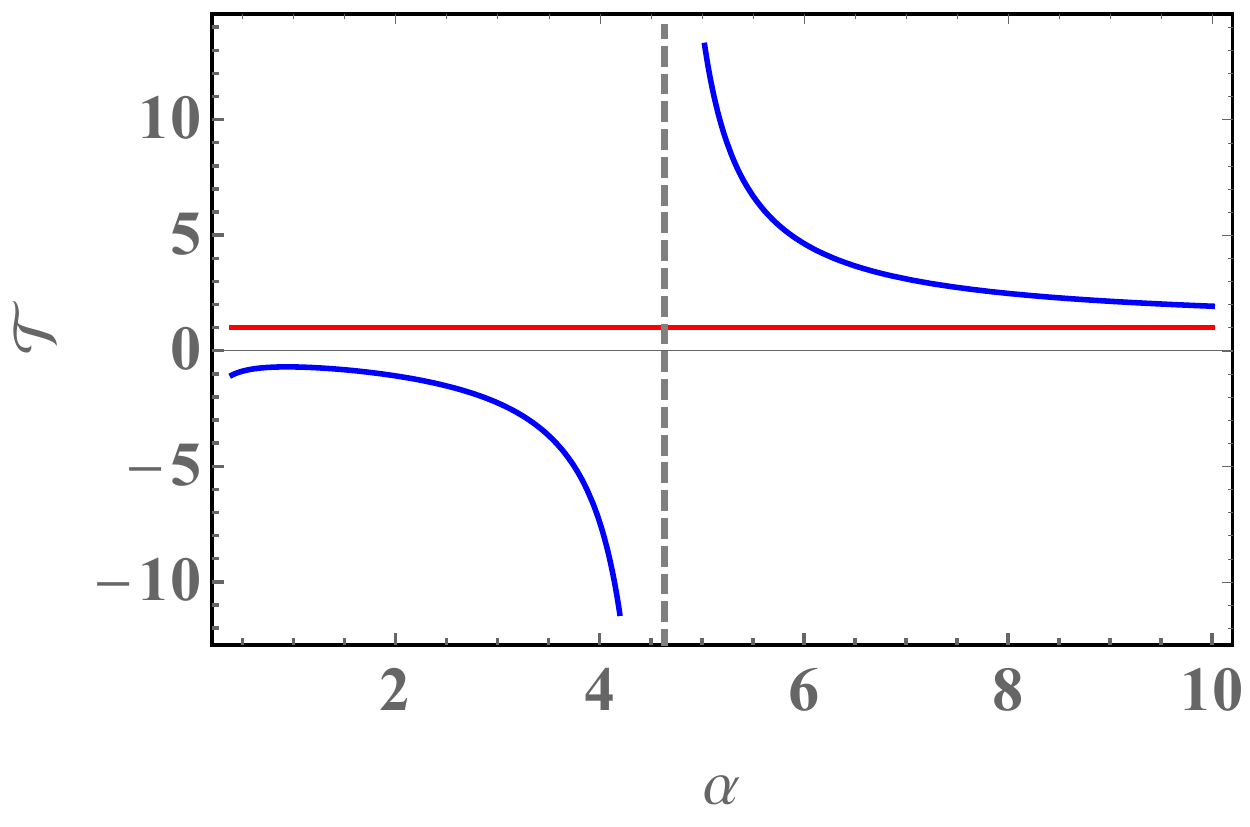}
\includegraphics[width=0.5\columnwidth]{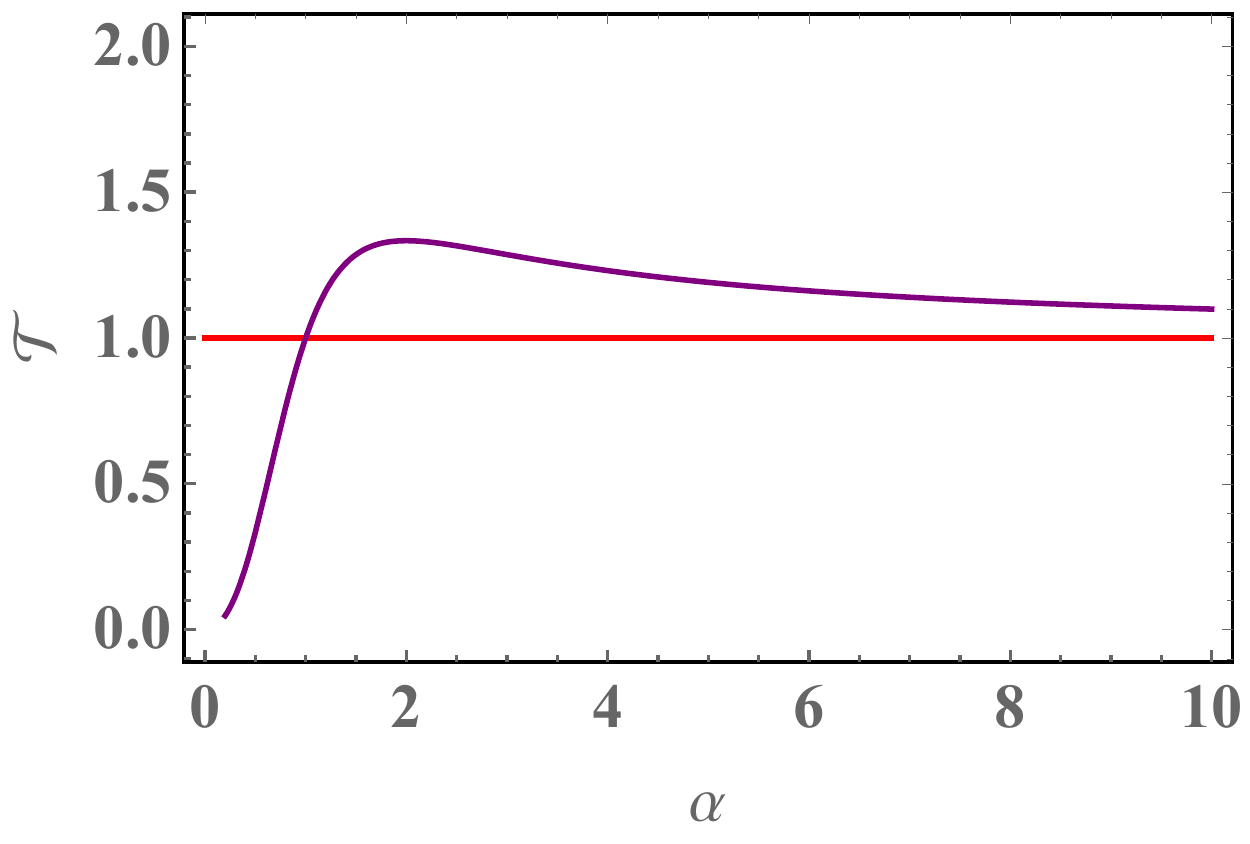}
\end{centering}
\caption{Effective temperatures for the underdamped case versus $\alpha $. Left Panel: values for $\mathcal{T}_{\rm i}$ are indicated by blue line plotted against constant $\mathcal{T}_{\rm k}$ represented by red line. The vertical line corresponds to the critical value $\alpha^{*} = 4.611 \ldots$. Right Panel: values for $\mathcal{T}_{\rm r}$ are indicated purple line  plotted against constant $\mathcal{T}_{\rm k}$ represented by red line. Both plots are with $m =  k = \gamma = 1 $.  and the kinetic temperature, $\mathcal{T}_{\rm k}$, is always kept constant at $ 1 $
}
\label{3-temperaturas} 
\end{figure}
From Eq.~(\ref{Tr-Tv}) and Eq.~(\ref{Ti-Tv}), we confirm the limit $\alpha \rightarrow \infty$, with $\mathcal{T}_{\rm k}$ fixed, attains the convergence of all effective temperatures to $\mathcal{T}_{\rm k}$, a unified effective measure. 
 Moreover, a comparison between $\mathcal{T}_{\rm i} $ and $\mathcal{T}_{\rm r}$ shows  the former is larger than the latter for $\alpha  > \alpha^{*}$, which implies a slower convergence to the  equilibrium-like case.
In Fig.~\ref{3-temperaturas}, we have depicted the three definitions of temperature we analysed. This account remains qualitatively the same whatever the values assumed for the parameters as long as the system remains underdamped.

\section{The overdamped limit}
  
The overdamped version of the previous model is defined by the limit $m/\gamma$ going to zero:
\begin{equation}
\frac{dx\left(t\right)}{dt}+ \frac{k}{\gamma} x(t) =  \frac{1}{\gamma} \, \zeta_{t}
\label{overdampedeq}
\end{equation}
Applying systematically the Fourier-Laplace method to Eq.~(\ref{overdampedeq}), we manage to obtain a general formula for the statistical moments of the position:
\begin{equation}
\avg{x^{2 n}}=\frac{a^{2 n} (2 n-1) !! }{k^n \prod _{j=1}^n (\alpha  \gamma +(2 j-1) k)}
\end{equation}
see appendix B for details on this calculation.

That formula provides us with the complete statistical information on the problem, enabling us to retrieve the probability distribution of the position  $p\left(x\right)$, by means of an inverse Fourier transform of the moment generating function, $\mathcal{G}_{x}\left(q\right)$:
\begin{equation}
p(x) = \int \exp \left[{\rm i} \, q \, x \right] \, \mathcal{G}_{x}\left(q\right) =  \int \exp \left[{\rm i} \, q \, x \right]  \sum_{n=0}^\infty \frac{{\rm i}^n \, q^n \, \left\langle x^{n} \right\rangle}{n!} dq
\end{equation}
from which we finally obtain \cite{gradshteyn}:
\begin{equation}
p(x)=\frac{k \, \Gamma \left[\frac{\alpha  \, \gamma }{2 \, k}+\frac{1}{2}\right] }{\sqrt{\pi } a \Gamma \left[\frac{\alpha  \gamma }{2 \, k}\right]}\left(1-\frac{k^2 }{a^2} x^2\right)^{\frac{\alpha  \, \gamma }{2 \, k}-1}
\label{pdf-position}
\end{equation}
Equation~(\ref{pdf-position}) concurs with a previous solution obtained by a field-theory approach~\cite{caceres}.

\subsection{Positional Temperature}

The average energy for the overdamped particle under dichotomous noise becomes: %
\begin{equation}
E = \frac{k}{2} \avg{x^2} =  \mathcal{T}_{\rm x}
\end{equation}
and therefrom we establish a positional temperature:
\begin{equation}
\mathcal{T}_{\rm x} = \frac{a^2}{2(k+\alpha \, \gamma  )}
\label{eod}
\end{equation}
which is the overdamped benchmark for the other effective temperatures, as $\mathcal{T}_{\rm k}$ is in the underdamped case.

We calculate the statistics of the asymptotic injective and dissipative energetic fluxes. 
Employing the Laplace-Fourier approach, we have obtained the average powers:

\begin{equation}
\avg{\it{j}_{\rm {inj}} } = - \avg{\it{j}_{\rm{dis}} } = 2 \, \alpha \, \mathcal{T}_{\rm x}
\end{equation}

\subsection{Intensive Temperature} 

Calculating the Gibbs entropy:
\begin{equation}
S =- \int p(x) \log p(x ) dx
\end{equation}
which for the steady state distribution~(\ref{pdf-position}) yields:
\begin{equation}
S = \log \left[\frac{\sqrt{\pi } \, a \Gamma \left[\frac{\alpha \gamma }{2 k }\right]}{k  \, \Gamma \left[\frac{\alpha \gamma +k }{2 k }\right]}\right]+
\frac{(\alpha \gamma -2 k ) \Gamma \left[\frac{\alpha \gamma }{2 k }+1\right] }{\alpha \gamma  \Gamma \left[\frac{\alpha \gamma }{2 k }\right]}
\left(\psi ^{(0)}\left[\frac{\alpha \gamma}{2 k }\right]-\psi ^{(0)}\left[\frac{\alpha \gamma +k }{2 k }\right]\right)
\label{entropyoverdamped}
\end{equation}
where $\psi ^{(0)}$ represents the Digamma function. 
The expression above may also be written as:
\begin{equation}
S = \log[E] + 
\log \left[\frac{ 2(\alpha  \gamma +k) \sqrt{ \pi} \, \Gamma \left[\frac{\alpha \gamma }{2 k }\right]}{a \, k  \, \Gamma \left[\frac{\alpha \gamma +k }{2 k }\right]}\right]+
\frac{(\alpha \gamma -2 k ) \Gamma \left[\frac{\alpha \gamma }{2 k }+1\right] }{\alpha \gamma  \Gamma \left[\frac{\alpha \gamma }{2 k }\right]}
\left(\psi ^{(0)}\left[\frac{\alpha \gamma}{2 k }\right]-\psi ^{(0)}\left[\frac{\alpha \gamma +k }{2 k }\right]\right)
\end{equation}
As stated earlier, the analytical formula for the system entropy provides us with the chance of developing further insight into the ansatz~(\ref{entropyansatz}) used in the underdamped case; notice the logarithmic dependence between Energy and Entropy. Expanding this formula up to the second power of $\alpha$ we get:

\begin{equation}
S =\log [E]+\frac{2 k}{\alpha  \gamma }+\log \left[\frac{ k }{a \, \alpha  \gamma e}\right]+\alpha \frac{  \gamma  \left(\pi ^2+6 \log [4]\right)}{6 k}-\alpha ^2 \frac{ \gamma ^2 \left(12 \, Z_r [3]+\pi ^2\right)}{8 k^2}+O\left(\alpha ^3\right)
\label{sodexpansion}
\end{equation}
Where $Z_r $ represents the Riemann Zeta function.
Taking the expression for the stationary mean energy in Eq.~(\ref{eod}) and the definition $h_{\alpha} = \left( h^{-1} + \frac{h^{\prime} }{\alpha^2} \right)^{-1}$ into consideration, we use the same expansion up until the second power in  Eq.~(\ref{entropyansatz}),  the ansatz $S= \log ( h_{\alpha} \, E )$, to get: 
\begin{equation}
S = \log[E]+2 \log \left[\alpha  \right]-\log[h']-\frac{\alpha ^2}{h \, h'}+O\left(\alpha ^3\right)
\label{sodexpansion2}
\end{equation}
In fact, Eq.~(\ref{sodexpansion2}) works as an approximation for Eq.~(\ref{sodexpansion}). Using both equations, it is possible to find the analytical forms of the terms $h$ and $h'$. However, both expressions shall be irrelevant to the relation between the intensive and kinetic temperatures in the overdamped regime, as we shall see further ahead.

\begin{figure}[tbh]
\begin{center}
\includegraphics[width=0.48\columnwidth]{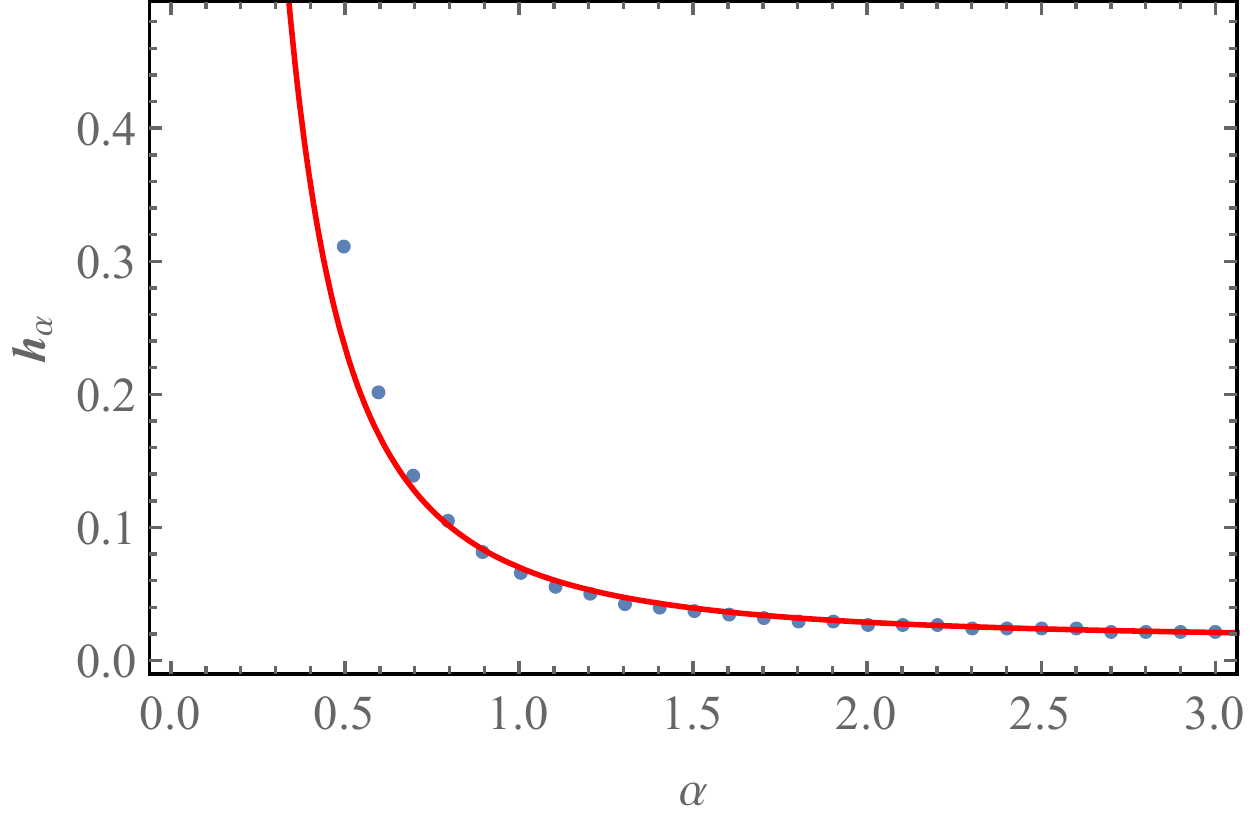} %
\includegraphics[width=0.48\columnwidth]{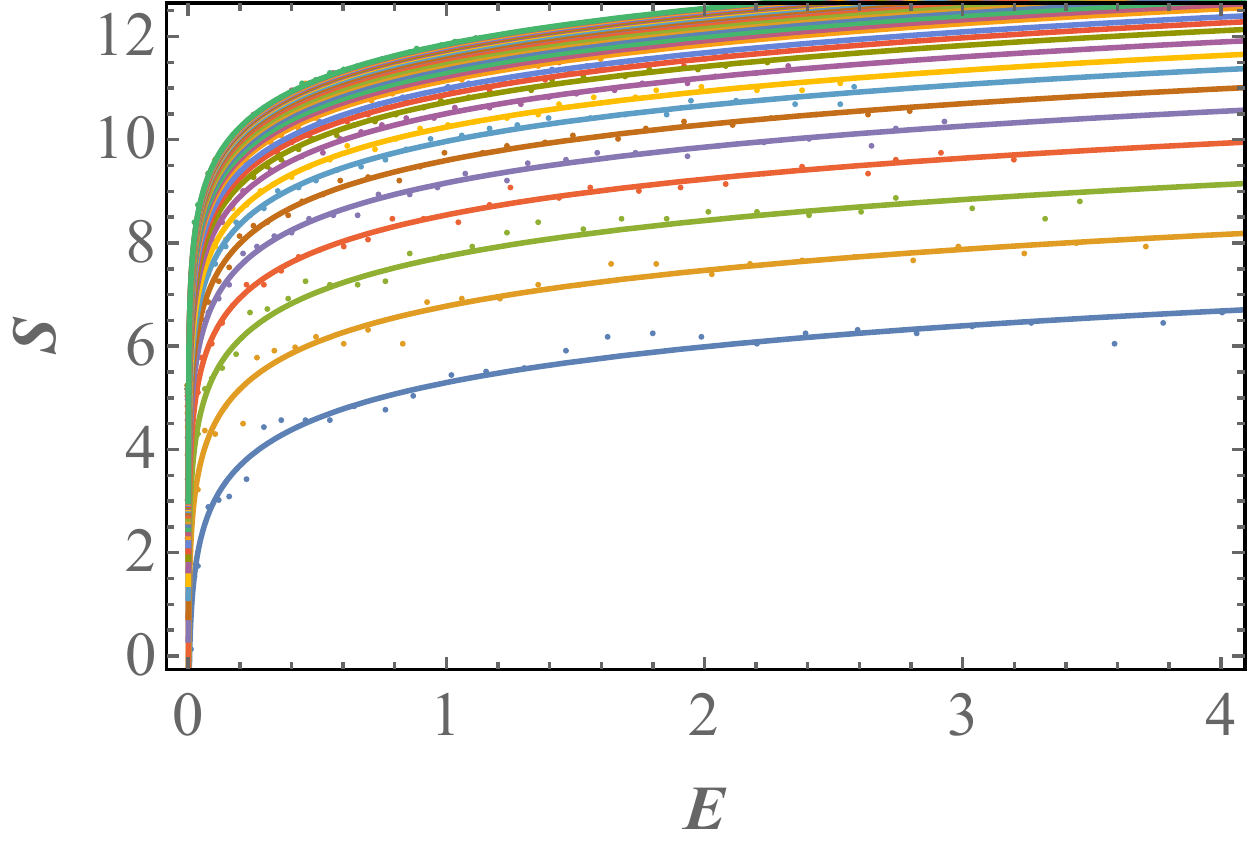} %
\end{center}
\caption{Left panel: plot for $h_\alpha$ in the overdamped case for mechanical parameters all equal to one; Right panel
Exponential fit for the relation between Energy and Entropy, for $\alpha$ and $a$ $\in$ $\{1/10,2/10,...,3\}$ and mechanical parameters equal to one
}
\label{residuos1}
\end{figure}

If we recast Eq.~(\ref{tibimodal}) for the overdamped case:
\begin{equation}
\label{eqn:SEanailitico}
\frac{dS}{dE}  \bigg\vert_{\mathcal{T}_{\rm x}} =    
\frac{\partial S}{\partial E}   \bigg\vert_{\alpha} +  \frac{\partial S}{\partial \alpha } \bigg\vert_{E}   \frac{\partial \alpha}{\partial E}  \bigg\vert_{\mathcal{T}_{\rm x}}  
\end{equation}
Since we have $E = \mathcal{T}_{\rm x} $;
\begin{equation}
 \frac{\partial \alpha}{\partial E}  \bigg\vert_{\mathcal{T}_{\rm x}}  =0
\end{equation}
We get:
\begin{equation}
\frac{dS}{dE}  \bigg\vert_{\mathcal{T}_{\rm x}} =    
\frac{\partial S}{\partial E}   \bigg\vert_{\alpha} = \frac{1}{E}
\end{equation}
That implies:
\begin{equation}
 \frac{dS}{dE}  \bigg\vert_{\mathcal{T}_{\rm x}} = \frac{1}{\mathcal{T}^{\rm o}_{\rm i}} = \frac{1}{E}
\end{equation}
and finally:
\begin{equation}
\mathcal{T}^{\rm o}_{\rm i} =  \mathcal{T}_{\rm x}
\label{Toi-Tx}
\end{equation}

\subsection{Response Temperature}

In the overdamped regime, the correlation for position yields:
\begin{equation}
\avg{x(t) x(t+s)} = a^2 \frac{k e^{- \alpha \,  s}-\alpha  \gamma  e^{-\frac{k \, s}{\gamma }}  }{k (k-\alpha  \, \gamma ) (k + \alpha \, \gamma )}
\label{corx-over}
\end{equation}
whereas the time-dependent average position reads:
\begin{equation}
\avg{x(t)}  =  \frac{a}{k-\alpha \, \gamma }\left(e^{- \alpha \, t} - e^{-\frac{k \, t}{\gamma }}\right)
\end{equation}
If we calculate the response to a constant field applied over the Hamiltonian at time $t_0$, we get:
\begin{equation}
\avg{x(t)}  =  \frac{a}{k-\alpha  \gamma } \left(e^{- \alpha \,  t} -e^{-\frac{k \, t}{\gamma }}\right)+\frac{f }{k} \, \left(e^{-\frac{k (t-t_0)}{\gamma }}-1\right) \, \Theta (t-t_0)
\end{equation}
which means that, integrating the response from $t_0$ until $t$, given that $s = t-t_0$, we obtain:
\begin{equation}
\chi(t+s,t) =   \frac{1}{k} \left(e^{-\frac{k \, s}{\gamma }}-1\right)
\label{sucep-over}
\end{equation}

Approaching the effective susceptibility $\chi_{eff}(s)$ and the corresponding response temperature $\mathcal{T}_{\rm r}$  with the same device as was employed for the underdamped case (see Eq.~(\ref{Tr})), we identify their corresponding separable factors in the position auto-correlation Eq.~(\ref{corx-over}) after comparing the white-noise limit with the actual susceptibility Eq.~(\ref{sucep-over}), see appendix D for step by step derivation.
\begin{equation}
C_x(s)  = \frac{a^2}{ 2 ( k + \alpha \gamma ) } \frac{ 2 \left(k e^{-\alpha s }-\alpha \,\gamma \, e^{-k \, s} \right)}{ k (k - \alpha \gamma)} = \mathcal{T}^{\rm o}_{\rm r} \chi_{eff}(s)
 \end{equation}
which gives us:
\begin{equation}
\chi_{eff}(s) =  \frac{ 2 \left(k e^{-\alpha s }-\alpha \,\gamma \, e^{-k \, s} \right)}{ k(k - \alpha \gamma)}
 \end{equation}

That said, in the overdamped case, the response temperature is equal to the positional temperature:
 \begin{equation}
 \mathcal{T}^{\rm o}_{\rm r} = \mathcal{T}_{\rm x}  =   \frac{a^2}{ 2(k + \alpha \gamma)}
 \label{Tor-tx}
 \end{equation}
Remarkably, we learn this approach provides us with the equivalence between the effective intensive and response temperatures, $\mathcal{T}^{\rm  \,o}_{\rm i} $ and $\mathcal{T}^{\rm \,o}_{\rm r}$, both equal to the positional temperature $\mathcal{T}_{\rm x}$, that is, $ \, \mathcal{T}^{\rm  \,o}_{\rm i} =  \mathcal{T}_{\rm x} = \mathcal{T}^{\rm  \,o}_{\rm r}  $.
 Given that the conditions of thermal equilibrium are not met within the framework of the overdamped particle subjected to the dichotomous noise, it is revealing of such a system that, engaging in the present analysis of effective temperatures, one is able to obtain a measure of effective temperature that encompasses three of the fundamental aspects of the canonical temperature.

\bigskip

\section{Concluding remarks}

Herein we have carried out an analysis over a set of definitions of effective temperature for a Generalised Langevin massive system in contact with dichotomous noise. Our work presents a quantitative description of three fundamental aspects of temperature that occur in non-equilibrium systems, and we have identified the asymptotic limit where the dichotomous noise mimics the features of a Langevin Gaussian reservoir, attaining the convergence of the three measures $\mathcal{T}_{\rm k}$, $\mathcal{T}_{\rm i}$ and $\mathcal{T}_{\rm r}$ for our model. We have demonstrated that it is possible to deepen the understanding of fundamental aspects with which effective temperature manifests its properties in the context of non-equilibrium.

By developing a quasi-canonical ensemble approach, through which we have compared different samples of the system with the fixed benchmark kinetic temperature $\mathcal{T}_{\rm k}$, we have presented a non-equilibrium generalization of the traditional canonical ensemble, which, in turn, assumes the fixed equilibrium temperature $T$. This framework offers a simple foundation for the development of the one to one relations between the different effective measures.

 We assert that the analogous of the temperature shall be exactly established by this collection of values, each representing a thermostatistical facet of the problem: its mean kinetic energy and the energetic fluxes ($\mathcal{T}_{\rm k}$), the response of system energy to changes in the information contained in it ($\mathcal{T}_{\rm i}$), and the effective response of the system associated with its auto-correlation ($\mathcal{T}_{\rm r}$). The same equivalence between all the effective temperatures is also achieved in the overdamped regime, which comes to indicate approach to equilibrium may be enforced upon the system by overdamping. It is also crucial to state that our results demonstrate there are alternative ways to measure the approach to equilibrium. Rather than solely applying the Kullback-Leibler divergence to quantify distance between the studied histograms and the Maxwell distribution, the comparison between the measures $\mathcal{T}_{\rm k}$, $\mathcal{T}_{\rm r}$ and $\mathcal{T}_{\rm i}$ also presents a relevant metric for the advance towards equilibrium, given that, without their convergence, the system is bound to be in non-equilibrium state.

For most of the underdamped case domain, our results have shown a system dependent hierarchy in the values of the three proposals, $\mathcal{T}_{\rm k} < \mathcal{T}_{\rm r} < \mathcal{T}_{\rm i}$ with all of them converging to the same value as we approach the features of a Gaussian reservoir. We ascribe this ordering to the exotic probability distributions associated with position and velocity of the massive particle under dichotomous noise
~\cite{JRM-SMDQ-2015}. The response and correlation functions, by contrast, basically preserve their usual Langevin functional forms, even though they are modified by the colour of the noise. Nevertheless, we have shown that $\mathcal{T}_{\rm i}$ can be negative --- i.e. the increase in entropy of the system does not come with an augment of its energy  --- particularly when the impact of the reservoir in the system can be solved by means of quasi-deterministic approaches. Along with the dichotomous noise case we have treated, we reckon systems subjected to  very rarefied granular gas reservoirs yield $\mathcal{T}_{\rm i} < 0$ as well. We have also identified a critical point in parameter space where $\mathcal{T}_{\rm i}$ becomes undefined, and the system completely loses the mutual relationship between energy and entropy, hence, at this particular set of the noise parameters, the dichotomous noise may render the system mean energy impervious to changes in the configuration space. 

Furthermore, Eq.~(\ref{modelo}) comprises an overwhelming range of experimental setups~\cite{ciliberto}, considering previous work on non-equilibrium reservoirs~\cite{br-jp}, we anticipate the differences between the various effective temperatures will be emphasised when the confining potential exhibits non-linearities, since their presence is likely to activate the higher-order cumulants of the dichotomous noise $\zeta_t$ making them act as supplementary sources of energy, as the cumulants trickle down to the second moment where energy is stored. We expect to further explore this question in future research.

In mentioning changes of information and energy of the system, we recall the performance of work and the cornerstone concept of absolute temperature of a system from which efficiency is computed. That said, it is worth enlarging the temperature set to a quantity, $\mathcal{T}_a$, as was done in Ref.~\cite{tact}, related to the efficiency of a system performing a Carnot-like cycle of thermostatiscal transformations along the lines implemented in Ref.~\cite{kurizki} for analysing a particular quantum Otto cycle. When dealing with a system in thermal equilibrium, all these quantities must collapse onto a single one for which it is possible to prove they are not only numerically equal, but to learn a conceptual one-to-one relation between the different definitions.

Besides, on systems described by the generalised Langevin Equation~(\ref{modelo}) we have aforementioned, similar surveys can be carried out on non-equilibrium systems to which proxies for the temperature were or can be considered, namely granular-matter systems~\cite{br-jp},  fluid activated dynamics~\cite{activated}, turbulence~\cite{turbulence}, simple fluids with preferred spatial direction (shear flow)~\cite{belousov-cohen}, interacting vortices in disordered superconductors~\cite{evaldo}, gene network models~\cite{gene}, biological self-assembly processes~\cite{self-assembly} and ratchets~\cite{ratchets} or even in finance, where the volatility of a financial instrument can be reinterpreted as an effective temperature as well~\cite{volatility}.

\begin{ack}
We would like to acknowledge P.~Maass and W.~A.~M.~Morgado for comments and discussions on this subject-matter as well as the Brazilian agencies CAPES and CNPq for financial support.
\end{ack}

\bigskip

\bigskip
\bigskip

\appendix

\section{The dichotomous noise}

Consider a stochastic process $\left\{ \zeta _{t}\right\} $ that
assumes two values $\zeta =\left\{ a,b\right\} $. The noise
switches from one state to the other in time according to pre-established 
rates:

\begin{itemize}
\item $\mu $ when going from $b$ to $a$;

\item $\bar{\mu}=\rho \,\mu $ otherwise.
\end{itemize}
That yields the following Master Equation,
\begin{equation}
\left\{
\begin{array}{ccc}
\frac{\partial \,p\left( a,t\,|\,\zeta _{0},t_{0}\right) }{\partial \,t} & =
& \mu \,p\left( b,t\,|\,\zeta _{0},t_{0}\right) -\bar{\mu}\,p\left(
a,t\,|\,\zeta _{0},t_{0}\right) \\
&  &  \\
\frac{\partial \,p\left( b,t\,|\,\zeta _{0},t_{0}\right) }{\partial \,t} & =
& \bar{\mu}\,p\left( a,t\,|\,\zeta _{0},t_{0}\right) -\mu \,p\left(
b,t\,|\,\zeta _{0},t_{0}\right)%
\end{array}%
\right. ,  \label{mastertelegraph}
\end{equation} 
where $p\left( \zeta_t ,t\,|\,\zeta _{0},t_{0}\right)$ is the transition probability for the stochastic variable $\zeta$ and $p\left( a,t\,|\,\zeta _{0},t_{0}\right) +p\left( b,t\,|\,\zeta
_{0},t_{0}\right) =1$. Assuming always the same initial condition,%
\begin{equation}
p\left( \zeta ,t_{0}\,|\,\zeta _{0},t_{0}\right) =\delta _{\zeta ,\zeta
_{0}}\,
\end{equation}
the solution to Eq.~(\ref{mastertelegraph}) reads,
\begin{equation}
\left\{
\begin{array}{ccc}
p\left( a,t\,|\,\zeta _{0},t_{0}\right) & = & r+\exp \left[ -\alpha \,\left(
t-t_{0}\right) \right] \,\left( \bar{r}\,\delta _{a,\zeta _{0}}-r\,\,\delta
_{b,\zeta _{0}}\right) \\
&  &  \\
p\left( b,t\,|\,\zeta _{0},t_{0}\right) & = & \bar{r}-\bar{r}\,\exp \left[
-\alpha \,\left( t-t_{0}\right) \right] \left( \bar{r}\,\delta _{a,\zeta
_{0}}-r\,\,\delta _{b,\zeta _{0}}\right)%
\end{array}%
\right.   \label{solucoes-telegraph}
\end{equation}

With $\alpha = \mu+\bar{\mu}$; for long enough time intervals, \textit{i.e.}, $t-t_{0}\gg \alpha ^{-1}$,
the stationary distribution reads,
\begin{equation}
p_{\mathrm{est}}\left( \zeta \right) =r\,\delta _{a,\zeta }+\bar{r}\,\delta
_{b,\zeta },
\end{equation}
where,
\begin{equation}
r\equiv \frac{\mu }{\mu +\bar{\mu}}=\frac{\mu }{\mu \,\left( 1+\rho \right) }%
=\frac{\mu }{\mu \,\hat{\rho}}=\hat{\rho}^{-1},\qquad \bar{r}\equiv 1-r=%
\frac{\hat{\rho}-1}{\hat{\rho}}
\label{f-funcao-mu}
\end{equation}

For the purpose of this work we have opted for simplicity to deal with the symmetric case $b = - a$ and $\rho = 1$.

From Eq.~(\ref{solucoes-telegraph}), one determines the evolution of the
noise between to instants $t$ and $t^{\prime }$,%
\begin{equation}
\begin{array}{ccc}
p\left( a,t\,|\,a,t^{\prime }\right) =r+\bar{r}\,\mathrm{e}^{-\alpha \left(
t-t^{\prime }\right) }, &  & p\left( a,t\,|\,b,t^{\prime }\right) =r-r\,%
\mathrm{e}^{-\alpha \left( t-t^{\prime }\right) }, \\
&  &  \\
p\left( b,t\,|\,a,t^{\prime }\right) =\bar{r}-\bar{r}\,\mathrm{e}^{-\alpha
\left( t-t^{\prime }\right) } ,&  & p\left( b,t\,|\,b,t^{\prime }\right) =%
\bar{r}+r\,\mathrm{e}^{-\alpha \left( t-t^{\prime }\right) }%
\end{array}%
\end{equation}%
\bigskip

Still, from the Master Equation, it is possible to determine the moments of $n$-th order of telegraph noise,
\begin{gather}
\left\langle \zeta \left( t_{1}\right) \,\ldots \,\zeta \left( t_{n}\right)
\right\rangle = 
\sum_{ \substack{\zeta \left( t_{1}\right) \,,\ldots \,,\zeta \left(
t_{n}\right) \\ \left( t_{1}>\ldots >t_{n}\right)} }
\left[ \zeta \left( t_{1}\right) \,\ldots \,\zeta
\left( t_{n}\right) \right] \,p\left( \zeta \left( t_{1}\right) ,\,\ldots ,\,\zeta
\left( t_{n}\right) \right)   \\      
= \sum_{\zeta \left( t_{1}\right) \,,\ldots \,,\zeta \left( t_{n}\right) }
\left[ \zeta \left( t_{1}\right) \,\ldots \,\zeta \left( t_{n}\right) \right]
\,p\left[ \zeta \left( t_{1}\right) |\zeta \left( t_{2}\right) \right]
\ldots p\left[ \zeta \left( t_{n-1}\right) |\zeta \left( t_{n}\right) \right]
\,p\left[ \zeta \left( t_{n}\right) \right]  \label{prod}
\end{gather}
Notice that when computing the general expressions for the moments of $n$-th order we must include all time orderings.

The general form of  the $n$-th order moments of $\zeta$ are quite intricate, but taking into consideration the symmetric properties of the telegraph noise assumed in this work the $2n$ order moment of $\zeta$ is equal to
\begin{equation}
\left\langle \zeta\left(t_{1}\right)\dots\zeta\left(t_{2n}\right)\right\rangle =a^{2n}\,\exp\left[-\alpha\sum_{l=1}^{n}\left(t_{2l-1}-t_{2l}\right)\right],\qquad\left(t_{1}>\dots>t_{2n}\right)
\label{prod2}
\end{equation}

Notice that reservoir symmetry implies that, given the ordering $\left(t_{1}>\dots>t_{2n}\right)$, only the correspondingly ordered product of transition probabilities in Eq.~(\ref{prod})  will give rise to non-vanishing contributions to Eq.~(\ref{prod2}).

To implement the numerical calculations of our analysis we have simulated the dichotomous noise $\zeta $ based on its first passage process. This means that we start by defining an initial value $\zeta_0$ as an initial condition, then, the solution for each possible $\zeta_0$ implies that the probability the noise changes from one value to another after an interval of time $\Delta t=t-t_{0}$ is equal to:
\begin{equation}
Q_{a\rightarrow -a }\left( \Delta t\right) = Q_{- a\rightarrow a }\left( \Delta t\right) = \mu \,\exp \left[ -\mu
\,\,\Delta _{t}\right]   \label{pba}
\end{equation}

If the noise is equal to $a\left( -a\right) $ at instant $t_{0}$, we pick a
number, $\Delta t$, exponentially distributed with characteristic scale $ \mu ^{-1}$; the noise keeps its value $a\left(
-a \right) $ and at instant $t_{0}+\Delta t=t$ it assumes a new value $-a \left( a\right) $. Then, for computational effects $t$ turns into the new
initial time and a new waiting time $\Delta t $ is computed and so forth.

\section{Laplace-Fourier Method}

Let us define the Laplace-Fourier transform as,
\begin{equation}
\widetilde{\mathcal{O}}\left(\mathrm{i}\,q+\varepsilon\right)\equiv\lim_{\varepsilon\rightarrow0}\int_{-\infty}^{\infty} \mathcal{O}\left(t\right)\,\mathrm{e}^{-\left(\mathrm{i}\,q+\varepsilon\right)\,t}\,dt.\label{laplace-fourier}
\end{equation}

Choosing the initial state $v(0)=0$ and $x(0)=0$ we obtain the following Laplace-Fourier transform of Eq.~(\ref{modelo}) 
\begin{equation}
\left\{ \begin{array}{ccc}
m\,\left(\mathrm{i}\,q+\varepsilon\right)\,\tilde{v}\left(\mathrm{i}\,q+\varepsilon\right) & = & -\gamma\,\tilde{v}\left(\mathrm{i}\,q+\varepsilon\right)-k\,\tilde{x}\left(\mathrm{i}\,q+\varepsilon\right)+\tilde{\zeta}\left(\mathrm{i}\,q+\varepsilon\right)\\
\\
\tilde{v}\left(\mathrm{i}\,q+\varepsilon\right) & = & \left(\mathrm{i}\,q+\varepsilon\right)\,\tilde{x}\left(\mathrm{i}\,q+\varepsilon\right)
\end{array}\right.
\end{equation}
Plugging the second line into the first one, we eliminate the velocity and we have for the position in reciprocal space 
\begin{equation}
\tilde{x}\left(\mathrm{i}\,q+\varepsilon\right)=\frac{\tilde{\zeta}\left(\mathrm{i}\,q+\varepsilon\right)}{R\left(\mathrm{i}\,q+\varepsilon\right)}
\end{equation}
The function $R\left(s\right)$ reads
\begin{equation}
R\left(s\right)=m\left(s-\kappa_{+}\right)\left(s-\kappa_{-}\right)
\end{equation}
with its zeros located at,
\begin{equation}
\kappa_{\pm}  =  -\frac{\theta}{2}\pm\mathrm{i}\,\sqrt{4\,\omega^{2}-\theta^{2}}
\end{equation}
where,
\begin{equation}
\theta=\frac{\gamma}{m},\qquad\omega^{2}=\frac{k}{m}
\end{equation}

That said, one learns that a generic quantity, $\mathcal{O}\left(t\right)$, is recast in reciprocal space as
\begin{equation}
\widetilde{\mathcal{O}}\left(\mathrm{i}\,q_{1}+\varepsilon\right)=h\left(\mathrm{i}\,q_{1}+\varepsilon\right)\,\tilde{x}\left(\mathrm{i}\,q_{1}+\varepsilon\right)\label{definition-O}
\end{equation}
For instance, for the velocity $h_{v}\left(s\right)=s$ whereas for the noise $h_{\zeta}\left(s\right)=R\left(s\right)$.

Since the system reaches a stationary state the soft ergodic property:
\begin{equation}
\left\langle \mathcal{O}\right\rangle \mathcal{=}\overline{\mathcal{O}}\equiv\lim_{\Xi\rightarrow\infty}\frac{1}{\Xi}\int_{0}^{\Xi} \mathcal{O}\left(t\right)\,\,dt\label{ergodic}
\end{equation}
 Eq.~(\ref{ergodic}) relates averages over samples, $\left\langle \mathcal{O}\right\rangle $, and averages over time, $\overline{\mathcal{O}}$, holds. Considering the final value theorem, we can connect the computation of statistics over time with the Laplace-Fourier transform~\cite{wamm-dosp}
\begin{equation}
\overline{\mathcal{O}}=\lim_{\Xi\rightarrow\infty}\frac{1}{\Xi}\int_{0}^{\Xi}\mathcal{O}\left(t\right)\,\,dt=\lim_{z\rightarrow0}\,z\int_{0}^{\infty}\,\mathrm{e}^{-z\,t}\,\mathcal{O}\left(t\right)\,dt\label{final-value-theo}
\end{equation}

\begin{figure}[tbh]
\begin{center}
\includegraphics[width=0.55\columnwidth,angle=0]{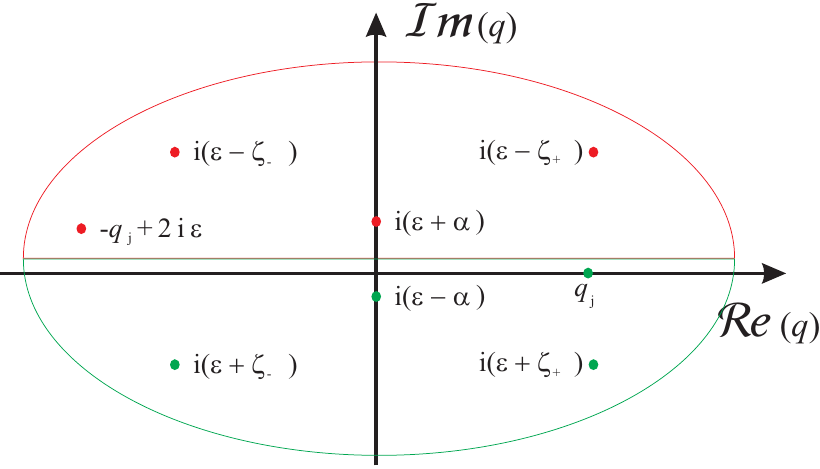} %
\end{center}
\caption{
Structure of the poles in Eq. (\ref{medias-final}) in the complex plane. The poles in the lower-arch will only be relevant in time dependent calculations such as the correlation function Eq.~(\ref{Cext}).
}
\label{residuos}
\end{figure}

Using Eq.~(\ref{laplace-fourier}) in Eq.~(\ref{final-value-theo}) as well as the equality between time and sample averaging in the stationary state we get for the $n$-th order moment
\begin{gather}
\left\langle \mathcal{O}^{n}\right\rangle 
\lim_{z\rightarrow0,\,\varepsilon\rightarrow0}\int\frac{z}{z-\sum_{l=1}^{n}\left(\mathrm{i}\,q_{l}+\varepsilon\right)}\frac{h\left(\mathrm{i}\,q_{1}+\varepsilon\right)\dots h\left(\mathrm{i}\,q_{n}+\varepsilon\right)}{R\left(\mathrm{i}\,q_{1}+\varepsilon\right)\dots R\left(\mathrm{i}\,q_{n}+\varepsilon\right)}\label{medias-final}\\
 \times\left\langle \tilde{\zeta}\left(\mathrm{i}\,q_{1}+\varepsilon\right)\dots\tilde{\zeta}\left(\mathrm{i}\,q_{n}+\varepsilon\right)\right\rangle \,\,\frac{dq_{1}}{2\pi}\dots\frac{dq_{n}}{2\pi}\nonumber
\end{gather}
The multiple integration in $q_{1},\dots,q_{n}$ eliminates all the modes related to the transient. Analytically, this means that only combinations of poles (see Fig.~\ref{residuos}) that lead to a final expression proportional to $z/z$ yield an \emph{a priori} non-vanishing solution. Only terms proportional to $\left[\sum_{l=1}^{\ell}\left(\mathrm{i}\,q_{l}+\varepsilon\right)\right]^{-1}$ are in agreement with that condition. From $R\left(q\right)$, those terms arise from the moments of the noise $\left\langle \tilde{\zeta}\left(\mathrm{i}\,q_{1}+\varepsilon\right)\dots\tilde{\zeta}\left(\mathrm{i}\,q_{n}+\varepsilon\right)\right\rangle $. 

Using the analytical result $\left\langle \zeta\left(t_{1}\right)\dots\zeta\left(t_{2n}\right)\right\rangle =a^{2n}\,\exp\left[-\alpha\sum_{l=1}^{n}\left(t_{2l-1}-t_{2l}\right)\right],$ with $\left(t_{1}>\dots>t_{2n}\right)$ for the noise moments we obtain;

\begin{equation}
\left\langle \tilde{\zeta}\left(\mathrm{i}\,q_{1}+\varepsilon\right)\dots\tilde{\zeta}\left(\mathrm{i}\,q_{2n}+\varepsilon\right)\right\rangle =\frac{a^{2n}}{\prod_{l=1}^{n}\left[\sum_{o=1}^{2l}\left(\mathrm{i}\,q_{o}+\varepsilon\right)\right]\prod_{l'=1}^{n}\left[\alpha+\sum_{o=1}^{2l'-1}\left(\mathrm{i}\,q_{o}+\varepsilon\right)\right]}
\label{corrtransf}
\end{equation}

\subsection{ Energetic Fluxes }

The direct application of Eq.~(\ref{medias-final}) for $n=2$ and $h(s) = s$ with Eq.~(\ref{corrtransf}) yields the kinetic temperature $\mathcal{T}_{\rm k}$ as presented in Eq.~(\ref{marconi}).

For the calculation of injected and dissipated powers we refer to Eq.~(\ref{Jdis}) and Eq.~(\ref{Jinj}). We get:

\begin{gather}
\it{j}_{\rm{inj}}^{\rm{(dichotomous)}} =\lim_{t\rightarrow \infty }\frac{1}{t}%
\int\limits_{0}^{t}\int dt^{\prime } \, \frac{dq_1}{2 \pi} \, \frac{dq_2}{2 \pi} \, \frac{%
e^{\left( \mathrm{i}\,q_{1}+\,%
\varepsilon \right) \,t' + \left( \mathrm{i}%
\,q_{2}+\varepsilon \right) \,t^{\prime }}
  \left( \mathrm{i}\,q_{1}+\varepsilon \right)}{ m^2\,R\left( \mathrm{i}%
\,q_{1}+\varepsilon \right)\,R\left( \mathrm{i}%
\,q_{2}+\varepsilon \right) }\left\langle \prod\limits_{l=1}^{2}\tilde{\zeta}%
\left( \mathrm{i}\,q_{l}+\varepsilon \right) \,\right\rangle   \label{JJinj}
\\
\it{j}_{\rm{dis}}^{\rm{(dichotomous)}} =- \gamma \lim_{t\rightarrow \infty }\frac{1}{t}%
\int\limits_{0}^{t}\int dt^{\prime } \, \frac{dq_1}{2 \pi} \, \frac{dq_2}{2 \pi} \, \frac{%
e^{\left( \mathrm{i}\,q_{1}+\,%
\varepsilon \right) \,t' + \left( \mathrm{i}%
\,q_{2}+\varepsilon \right) \,t^{\prime }}
   \left( \mathrm{i}\,q_{1}+\varepsilon \right)\left( \mathrm{i}\,q_{2}+\varepsilon \right)}{ m^2\,R\left( \mathrm{i}%
\,q_{1}+\varepsilon \right)\,R\left( \mathrm{i}%
\,q_{2}+\varepsilon \right) }\left\langle \prod\limits_{l=1}^{2}\tilde{\zeta}%
\left( \mathrm{i}\,q_{l}+\varepsilon \right) \,\right\rangle    \label{JJdis}
\end{gather}
from which we obtain the results presented in Eq.~(\ref{Jdis}) and Eq.~(\ref{Jinj}) .

\subsection{Underdamped Position Auto-Correlation}

The Laplace-Fourier method was also utilised to compute the long-term correlation functions:
\begin{equation}
\mathrm{C}_{\mathcal{O}}(s)\equiv \lim_{t\rightarrow \infty }\frac{1}{t}%
\int\limits_{0}^{t}\left\langle \mathcal{O}\left( t^{\prime }\right) \,\mathcal{O}\left(
t^{\prime }+s\right) \right\rangle \,dt^{\prime }
\label{cor-vel}
\end{equation}
which translates into,
\begin{equation}
\mathrm{C}_{\mathcal{O}}\left( s\right) =\lim_{t\rightarrow \infty }\frac{1}{t}%
\int\limits_{0}^{t}\int dt^{\prime }e^{\left( \mathrm{i}\,q_{2}+\,%
\varepsilon \right) \,s}\prod\limits_{j=1}^{2}\frac{dq_{j}}{2\pi }\frac{%
h \left( \mathrm{i}\,q_{j}+\varepsilon \right) e^{\left( \mathrm{i}%
\,q_{i}+\varepsilon \right) \,t^{\prime }}}{m\,R\left( \mathrm{i}%
\,q_{j}+\varepsilon \right) }\left\langle \prod\limits_{l=1}^{2}\tilde{\zeta}%
\left( \mathrm{i}\,q_{l}+\varepsilon \right) \,\right\rangle  \label{Cext}
\end{equation}

For the position, $h(.) = 1$, and expressing the noise terms as Eq.~(\ref{corrtransf}) indicates we obtain the correlation function of the position:
\begin{gather}
\mathrm{C}_{x}\left( s\right)  = 
\frac{a^2 }{2 \, m \, \gamma  \, k \Omega \, \left(\left(k+\alpha ^2 m\right)^2-\alpha ^2 \gamma ^2\right)} \{2\, m\, k \, \gamma\,  \Omega \, e^{- \alpha \, s }
\\ 
\label{correlation}
 +
\alpha \, e^{-\frac{\gamma  s}{2 m}} \left[\gamma  \left(-\gamma ^2+3 k m+\alpha ^2 m^2\right) \sin \left(\Omega \, s\right)+ 2\,m\,\Omega \left(m \left(k+\alpha ^2 m\right)-\gamma ^2\right) \cos \left(\Omega \, s\right) \right] \} 
\nonumber   
\end{gather}
\begin{figure}[tbh]
\begin{center}
\includegraphics[width=0.53\columnwidth,angle=0]{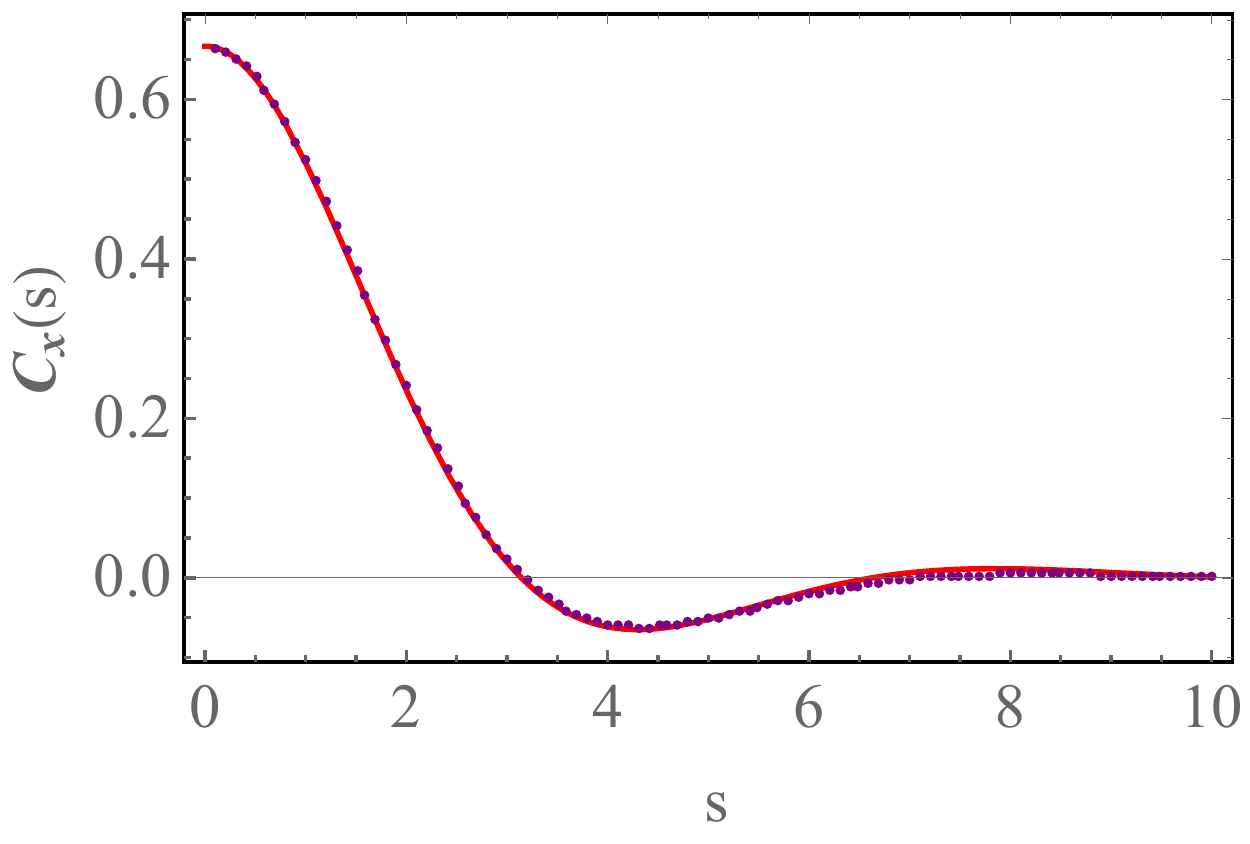} %
\end{center}
\caption{Correlation of the position versus lag. The dots are obtained from the numerical implementation of  Eq.~(\ref{modelo})  and the line corresponds to Eq.~(\ref{correlation}). All the parameters in this case have been set to $1$
}
\label{correlation-fig}
\end{figure}
Where $\Omega \equiv \frac{\sqrt{\gamma ^2-4 k m}}{2 m}$.
After redefining the factors and the coefficients in Eq.~(\ref{eqeqcorrelation}) and using $ \Lambda \equiv m \left(k+\alpha ^2 m\right)-\gamma ^2$  we obtain the response temperature $\mathcal{T}_{\rm r}$ in Eq.~(\ref{Tr-Tv}). The accurateness of Eq.~(\ref{eqeqcorrelation}) is verified in Fig.~\ref{correlation-fig}

\subsection{Overdamped Position moments}

The calculation for the overdamped moments of position $x$ is equally done by Laplace-Fourier transforming the model equation and subsequently applying the time averaging procedure by integrating in complex plane.

From Eq.~(\ref{overdampedeq}), we get $R(s) = \gamma (s - \kappa)$, with $\kappa = -\frac{k}{\gamma }$. The calculation of the position moments $\avg{x^n}$  becomes:

\begin{equation}
\avg{x^{2 n}} = \lim_{t\rightarrow \infty }\frac{1}{t}%
\int\limits_{0}^{t} dt^{\prime } \int  \, \prod_{i=1}^{2n} \frac{dq_i}{2 \pi} \, \frac{%
e^{\left( \mathrm{i}\,q_{i}+\,%
\varepsilon \right) \,t' }
 }{ \gamma \,R\left( \mathrm{i}%
\,q_{1}+\varepsilon \right)\, }\left\langle \prod\limits_{l=1}^{2n}\tilde{\zeta}%
\left( \mathrm{i}\,q_{l}+\varepsilon \right) \,\right\rangle 
\end{equation}

Note that all the odd moments will be zero given the symmetric quality of the model and the reservoir. The simplicity of the overdamped model allows us to easily obtain the first four even moments, which become:

\begin{gather}
\avg{x^2 }= \frac{a^2}{k (\alpha  \gamma +k)}
\\
\avg{x^4 }= \frac{3 a^4}{k^2 (\alpha  \gamma +k) (\alpha  \gamma +3 k)}
\\
\avg{x^6 }=\frac{15 a^6}{k^3 (\alpha  \gamma +k) (\alpha  \gamma +3 k) (\alpha  \gamma +5 k)}
\\
\avg{x^8 }= \frac{105 a^8}{k^4 (\alpha  \gamma +k) (\alpha  \gamma +3 k) (\alpha  \gamma +5 k) (\alpha  \gamma +7 k)}
\end{gather}

We then extrapolate these results to get the numerically verified formula:

\begin{equation}
\avg{x^{2 n}}=\frac{a^{2 n} (2 n-1) !! }{k^n \prod _{j=1}^n (\alpha  \gamma +(2 j-1) k)}
\end{equation}

\section{Intensive Temperature}

The entropy is a function of the energy of the system, $E$, and the reservoir colour parameter, $\alpha $,
\begin{equation}
\label{eqn:SEalpha}
dS = \frac{\partial S}{\partial E} \bigg\vert_{\alpha}  \, dE  +  \frac{\partial S}{\partial \alpha }  \bigg\vert_{E} \, d\alpha 
\end{equation}
In addition, we assume the following relations, given that the parameters $\alpha$ and $a$ exhaust the mathematical information contained in the reservoir, we may treat kinetic temperature $a$ or $\alpha$ as function of the other two variables associated with the reservoir%
\begin{equation}
\label{eqn:alphaAT}
d\alpha = \frac{\partial \alpha}{\partial a} \bigg\vert_{\mathcal{T}_{\rm k}} \, da   +  \frac{\partial \alpha}{\partial \mathcal{T}_{\rm k}}  \bigg\vert_{\alpha} \, d\mathcal{T}_{\rm k} 
\end{equation}
and;
\begin{equation}
\label{eqn:AETM}
da = \frac{\partial a}{\partial E} \bigg\vert_{\mathcal{T}_{\rm k}} \, dE   +  \frac{\partial a}{\partial \mathcal{T}_{\rm k}}  \bigg\vert_{E} \,d\mathcal{T}_{\rm k} 
\end{equation}
Plugging Eq.~(\ref{eqn:alphaAT}) into  Eq.~(\ref{eqn:SEalpha}) and dividing by $dE$ we obtain:
\begin{equation}
 \frac{dS}{dE}  =   \frac{\partial S}{\partial E}   \bigg\vert_{\alpha} +  \frac{1}{dE}  \frac{\partial S}{\partial \alpha }  \bigg\vert_{E}  \left( \frac{\partial \alpha}{\partial a} \bigg\vert_{\mathcal{T}_{\rm k}} \, da  +  \frac{\partial \alpha}{\partial \mathcal{T}_{\rm k}} \bigg\vert_{\alpha} d\mathcal{T}_{\rm k}  \right) \, 
\end{equation}
Since we have fixed the kinetic temperature, i.e., $d\mathcal{T}_{\rm k}=0$, we get:
\begin{equation}
 \frac{dS}{dE}  \bigg\vert_{\mathcal{T}_{\rm k}} =   \frac{\partial S}{\partial E}   \bigg\vert_{\alpha} + \frac{1}{dE}  \frac{\partial S}{\partial \alpha } \bigg\vert_{E} \left[ \frac{\partial \alpha}{\partial a} \left( \frac{\partial a }{\partial E} \bigg\vert_{\mathcal{T}_{\rm k}} \, dE  +  \frac{\partial a}{\partial \mathcal{T}_{\rm k}} \bigg\vert_{E} \, d\mathcal{T}_{\rm k} \right) \right]
\end{equation}
Using again condition $d \mathcal{T}_{\rm k} = 0$:
\begin{equation}
 \frac{dS}{dE}  \bigg\vert_{\mathcal{T}_{\rm k}} =    
  \frac{\partial S}{\partial E}   \bigg\vert_{\alpha} +  \frac{\partial S}{\partial \alpha } \bigg\vert_{E}   \frac{\partial \alpha}{\partial E}  \bigg\vert_{\mathcal{T}_{\rm k}}  
\end{equation}
Recalling the energy of the system reads 
\begin{equation}
E = \frac{a^2 (\gamma +2 \alpha  m)}{2 \gamma  \left(\alpha  \gamma +k+\alpha ^2 m\right)}
= \mathcal{T}_{\rm k} \, \left[1+\frac{ \gamma \, \mathcal{T}_{\rm k} }{2 \alpha m} \right] \end{equation}
we get,
\begin{equation}
\frac{\partial E}{\partial \alpha }\bigg\vert_{\mathcal{T}_{\rm k}} = - \frac{\gamma}{ 2 \,  \alpha^2 m} \mathcal{T}_{\rm k}
\end{equation}
and finally,
\begin{equation}
\frac{dS}{dE}  \bigg\vert_{\mathcal{T}_{\rm k}} =   \frac{\partial S}{\partial E}   \bigg\vert_{\alpha} -  \frac{2 \,  \alpha^2 m}{\gamma \,  \mathcal{T}_{\rm k}} \frac{\partial S}{\partial \alpha }  \bigg\vert_{E}
\end{equation}

\section{Calculating $\chi_{eff}$}

\subsection{Underdamped System}

The definition of the effective susceptibility $\chi_{eff}$, as stated in the main text, establishes an analogous relation to the FDT, and is based in the observation of changes that occurr to the position auto-correlation as the reservoir approaches the Gaussian limit $\alpha \rightarrow \infty$ with $\mathcal{T}_{\rm k}$ kept constant;

Taking the traditional Langevin model, where $\eta(t)$ is Gaussian noise with $\avg{\eta(t)} = 0$ and $\avg{\eta^2(t)} = 2 \gamma T$;

\begin{equation}
m\frac{d^{2}x\left(t\right)}{dt^{2}}=-\gamma\,\frac{dx\left(t\right)}{dt}-k\,x\left(t\right)+\eta_{t}, \quad v\left(t\right) \equiv \frac{dx\left(t\right)}{dt}
\label{langevinmodel}
\end{equation}

We may obtain the position auto-correlation function ($C_x(s) = \langle x(t) x(t+s)  \rangle$) by the same Laplace-Fourier approach we used for the dichotomous case to get the canonical result (see Risken chapter 7 \cite{risken}). Note the superscript $l$ is used to point out the formulas associated with the Langevin case;

\begin{equation}
C^l_x(s) = T \frac{e^{-\frac{\gamma  s}{2 m}}}{k}   \left(\cos \left[\frac{s}{2 m} \sqrt{4 k m-\gamma ^2}\right]-\frac{\gamma  }{\sqrt{4 k m-\gamma ^2}} \sin \left[\frac{s}{2 m} \sqrt{4 k m-\gamma ^2}\right] \right)
\label{langevincorrelation}
\end{equation}

Given that thermal equilibrium is reached, the fluctuation dissipation relation Eqs.~(\ref{eqresponse})(\ref{Tr}) is verified, we recall it:

\begin{gather}
\frac{ \partial C^l_x \left(s \right)}{\partial s } \equiv  T\, R \left(s \right)  
\\
C^l_x (0) - C^l_x(s) \equiv T \, \chi (s).
\end{gather}

Now we apply the Gaussian limit to the position auto-correlation in the model subjected to dichotomous noise and observe how the correlation changes until reaching the limit. The complete result, simplified for the sake of simplicity in Eq.~(\ref{eqeqcorrelation}), reads:
\begin{gather}
 C_x(s) =  \frac{1}{m k} \mathcal{T}_{\rm k} \frac{ \alpha ^2  m^2   + m k -\gamma ^2}{ \alpha^2   m    -\gamma \alpha  + k} 
\bigg(\frac{\gamma  k }{\alpha ^3 m^2+ \alpha k m- \alpha \gamma ^2} \, e^{- \alpha  s} + \nonumber
\\
+  e^{-\frac{\gamma  s}{2 m}} \left(\frac{\gamma  \left(\alpha ^2 m^2+3 k m -\gamma ^2\right) }{\sqrt{4 k m-\gamma ^2} \left(\alpha ^2 m^2+k m -\gamma ^2\right)} \sin \left[\frac{s}{2 m} \sqrt{4 k m-\gamma ^2}\right]+\cos \left[\frac{s}{2 m} \sqrt{4 k m-\gamma ^2}\right]\right)\bigg) 
\label{correlationcomplete}
\end{gather}
In the limit of $\alpha \rightarrow \infty$ the  decaying factor $e^{-\alpha s}$ associated to the reservoir persistence time $\alpha^{-1}$ goes to zero. That leaves us with:
\begin{gather}
\lim_{\alpha \rightarrow 0} C_x(s) =  \frac{1}{m k} \mathcal{T}_{\rm k} \frac{   \left( \alpha ^2 m^2 +(...)\right)}{   ( \alpha^2 m    +(...))} \nonumber
\\
 e^{-\frac{\gamma  s}{2 m}}  \left( \left(\frac{\alpha ^2 m^2+(...)}{\alpha ^2 m^2+(...)}\right)\frac{\gamma   }{\sqrt{4 k m-\gamma ^2} } \sin \left[\frac{s}{2 m} \sqrt{4 k m-\gamma ^2}\right]+\cos \left[\frac{s}{2 m} \sqrt{4 k m-\gamma ^2}\right]\right)
\end{gather}
which becomes, following l'H\^opital's rule:
\begin{gather}
\lim_{\substack{\alpha \rightarrow 0 \\ \mathcal{T}_{\rm k} = \, const}} C_x(s) =  \frac{1}{k} \mathcal{T}_{\rm k}
 e^{-\frac{\gamma  s}{2 m}}  \left( \frac{\gamma   }{\sqrt{4 k m-\gamma ^2} } \sin \left[\frac{s}{2 m} \sqrt{4 k m-\gamma ^2}\right]+\cos \left[\frac{s}{2 m} \sqrt{4 k m-\gamma ^2}\right]\right)
 \label{limcor}
\end{gather}

Comparing Eq.~(\ref{limcor}) and Eq.~(\ref{langevincorrelation}) we observe they are very similar except for a sign in the oscillatory terms. Note that all dependency in $\alpha$ is now stored at the kinetic temperature $\mathcal{T}_{\rm k}$, and the remaining expression is a slightly modified form of the linear susceptibility found both in the Langevin system and in our dichotomous model. Namely;

\begin{gather}
\chi^l(s) =  \frac{1}{T}(C_x^l (0) - C_x^l (s)) = \nonumber\\   = \frac{1}{k} \left( 1 -
 e^{-\frac{\gamma  s}{2 m}}  \left( \frac{-\gamma   }{\sqrt{4 k m-\gamma ^2} } \sin \left[\frac{s}{2 m} \sqrt{4 k m-\gamma ^2}\right]+\cos \left[\frac{s}{2 m} \sqrt{4 k m-\gamma ^2}\right]\right) \right)
\end{gather}

As the same mechanical traits are shared by both systems, the linear susceptibility of our dichotomous noise model to the action of a small steady constant field  has the same form as the expression above, and may also be evaluated by the Laplace-Fourier Method.

We conclude that there are separable factors in the original expression for the dichotomous position auto-correlation (Eq.~(\ref{eqeqcorrelation}) or Eq.~(\ref{correlationcomplete})) that  converge (in the white-noise limit), respectively, to $\mathcal{T}_{\rm k}$ and to the modified Linear susceptibility. The first one is none but the response temperature as defined in the main text:
\begin{equation}
 \mathcal{T}_{\rm r} = \mathcal{T}_{\rm k} \frac{ \alpha ^2  m   +  k -\gamma ^2/m}{ \alpha^2   m    -\gamma \alpha  + k} 
\end{equation}
The second one becomes the effective susceptibility:
\begin{gather}
\chi_{eff}(s) =  \frac{e^{- \alpha  s}}{ k}  \left(\frac{\gamma  k }{\alpha ^3 m^2+ \alpha k m- \alpha \gamma ^2}\right)  + \nonumber
\\
+  \frac{e^{-\frac{\gamma  s}{2 m}}}{k} \left(\frac{\gamma  \left(\alpha ^2 m^2+3 k m -\gamma ^2\right) }{\sqrt{4 k m-\gamma ^2} \left(\alpha ^2 m^2+k m -\gamma ^2\right)} \sin \left[\frac{s}{2 m} \sqrt{4 k m-\gamma ^2}\right]+\cos \left[\frac{s}{2 m} \sqrt{4 k m-\gamma ^2}\right]\right)
\end{gather}

\subsection{Overdamped System}

Using the same method as was described for the underdamped case, we now have the model:

\begin{equation}
\gamma\,\frac{dx\left(t\right)}{dt}= -k\,x\left(t\right)+\eta_{t}, \quad v\left(t\right) \equiv \frac{dx\left(t\right)}{dt}
\label{ODlangevinmodel}
\end{equation}
that yields the following position auto-correlation:
\begin{equation}
C^{l \rm o}_x(s) = \frac{T }{k} e^{-\frac{k s}{\gamma }}
\label{ODcorrlange}
\end{equation}

Note the application of a constant force implies in the following formula for susceptibility, already established in the main text as Eq.~(\ref{sucep-over}). This result is equivalent for both the overdamped Langevin case and our dichotomous overdamped model, as the two systems share the same mechanical traits.

\begin{equation}
\mathcal{X}(t,t+s) =   \frac{1}{k} \left(e^{-\frac{k \, s}{\gamma }}-1\right).
\end{equation}

The position auto-correlation for the overdamped particle in interaction with dichotomous noise, mentioned in the main text in Eq.~(\ref{corx-over}) is:

\begin{equation}
C^{\rm o}_x(s) = a^2 \frac{k e^{- \alpha \,  s}-\alpha  \gamma  e^{-\frac{k \, s}{\gamma }}  }{k (k-\alpha  \, \gamma ) (k + \alpha \, \gamma )}
\end{equation}
Using Eq.~(\ref{eod}) we may rewrite the expression above as:
\begin{equation}
C^{\rm o}_x(s) = \mathcal{T}_{\rm x} \frac{ 2 \left(k e^{-\alpha s }-\alpha \,\gamma \, e^{-k \, s} \right)}{ k (k - \alpha \gamma)}
\end{equation}
Applying the white-noise limit, defined as $\alpha \rightarrow 0$ with the positional temperature $\mathcal{T}_{\rm x}$ kept constant we get;
\begin{equation}
\lim_{\substack{\alpha \rightarrow 0 \\ \mathcal{T}_{\rm x} = \, const}} C^{\rm o}_x(s) =  \frac{2 \mathcal{T}_{\rm x} }{k} e^{-\frac{k s}{\gamma }}
\end{equation}
That leaves us with the effective susceptibility:
\begin{equation}
\chi_{eff}^{ \rm o}(s) =  \frac{2}{k} e^{-\frac{k s}{\gamma }}
\end{equation}
And the response temperature;:
\begin{equation}
\mathcal{T}^{\rm o}_r = \mathcal{T}_{\rm x}
\end{equation}

Note that this result for the overdamped particle is consistent with the one for the underdamped particle. 

\par The white-noise limit provides many equilibrium-like qualities to the system, such as the convergence of the three effective temperatures, the functional form of the average energy fluxes (through the use of kinetic/positional temperatures)\cite{JRM-SMDQ-2015}, and the Gaussian probability distribution functions for position and velocity\cite{SMDQ-FR}, but does not verify FDT.

{}

\end{document}